# KNOWLEDGE VALUE STREAM FRAMEWORK FOR COMPLEX PRODUCT DESIGN DECISIONS

Dr. Ramakrishnan Raman

Honeywell Technology Solutions Lab, Bangalore, India

ramakrishnan.raman@honeywell.com

Dr. Meenakshi D'Souza

International Institute of Information Technology Bangalore, India

meenakshi@iiitb.ac.in

***Abstract***

Product Development value stream includes all the activities (value added, and non-value added) for designing and developing a product. It is characterized by flow of knowledge that drives the decisions, and deals with how the product is conceptualized, architected and designed by the design teams. The intangible flow of knowledge is determined by knowledge value stream – which shapes how the raw concepts/ ideas flow into mature knowledge, how the knowledge is socialized, internalized, and how the knowledge impels the decisions in the product development value stream. For complex products, the design teams encounter tough challenges, such as uncertainty and variability, while making design decisions. This paper proposes a framework for knowledge value stream for complex product design decisions. The framework encompasses knowledge cadence and learning cycles as its core elements and incorporates rudiments for complex product design such as uncertainty, variability and perceptions. It helps in managing uncertainty and variability and evolving the required knowledge to drive optimal decisions during the design of complex products. It advocates a phased model for framework deployment towards establishing and progressively maturing the knowledge value stream.

# KNOWLEDGE VALUE STREAM FRAMEWORK FOR COMPLEX PRODUCT DESIGN DECISIONS

Ramakrishnan Raman
Honeywell Technology Solutions Lab, Bangalore
ramakrishnan.raman@honeywell.com

Meenakshi D'Souza
International Institute of Information Technology, Bangalore
meenakshi@iiitb.ac.in

## I. INTRODUCTION

Value creation drives successful innovation, and this value creation surfaces the need for design and development of great successful products that add value to customers and various relevant stakeholders. A value stream includes all activities that are necessary to create a product and make it available to the customer. From the perspective of value stream, the activities may be value added, non-value added (waste) and supporting nature (McManus, 2005). The manufacturing/ production value stream deals with how the product is assembled, manufactured, stored and sourced. The product development value stream deals with how the product is conceptualized, architected, designed and developed. One of the primal focus areas of lean principles is in ensuring efficient flow in the value stream (Reinertsen, 2009). Flow is said to be efficient when value creation happens unhampered continuously, with minimal wastes. Product development typically involves phases such as Concept Development, Requirements, Architecture, Design, Implementation and Testing/Verification. Most of the critical architectural and design decisions are made in the early phases of product design (in this paper, the terms "architecture" and "design" are considered synonymous from the viewpoint of decision making), and these decisions have a significant bearing on how the product turns out to be, how it is realized and used by the customer. Unlike in a manufacturing environment where "visible" material flows from raw material to finished product, it is the knowledge that flows often in an "invisible" manner in a product design environment. Knowledge Value Stream focuses on how the knowledge flows to drive the decisions across the product development value streams. Ideally, a successful product (a product that is released to the market at the right time, with high quality performance and standing ahead of competition) implies that the optimal right decisions were taken at the right point in time during the product design and development lifecycle. This implies an efficient and effective knowledge value stream, since it is the knowledge that impels the decisions. For complex products, the design teams face tough challenges in making the right design decisions, due to significant uncertainty, variability, and knowledge gaps. Wrong decisions often result in causing defects, rework loopbacks, cost overruns, product under-performance, and product falling behind competition.

Significant research has been undertaken on various aspects pertaining to knowledge value stream, including managing product development knowledge, knowledge wastes, knowledge based engineering, and addressing uncertainty in complex product design and development (Kennedy, 2003) (Morgan and Liker, 2006) (Cabello et. al., 2012) (Tyagi et.al., 2015) (Ringen and Welo, 2015). But, an integrated framework for knowledge value stream that strings together various aspects pertaining to knowledge flow, knowledge creation and management, in product design and development context, has not been adequately addressed. Typical wastes in knowledge value stream include those causing unclear/ revisited decisions, re-learning, confusion, incorrect understanding and longer feedback cycles. An optimal knowledge value stream, with reduced wastes, would ensure an efficient flow and evolution of knowledge across the organization, thereby driving the right and robust design decisions, and enabling successful products. This paper proposes a framework for knowledge value stream for complex product design and development, towards driving robust design decisions. The proposed framework encompasses knowledge cadence and learning cycles as its core elements, and incorporates aspects such as uncertainty, variability, closed feedback loops and perceptions. In addition, this paper outlines a phase-wise deployment approach for maturing the various elements of the proposed framework.

The rest of the paper is organized as follows: Section II discusses various aspects and the related work pertaining to complex product design and development, including prevalent knowledge, knowledge gaps, uncertainty and variability, learning cycles, and how these are addressed in product design and development. Section III discusses the proposed knowledge value stream framework, while Section IV outlines a deployment approach for the framework. Section V finally has the conclusions and future work.

## II. COMPLEX PRODUCT DESIGN & DEVELOPMENT

This section sets the context for complex product design and development, and discusses the related work pertaining to specific nuances such as knowledge gaps, uncertainty and variability in the design of complex products. Further, this section introduces the concepts of prevalent knowledge, learning cycles and its implications on product design.

### A. *Product Design & Development*

In product design and development, the design teams comprising product architects, designers, and engineers engage in knowledge intensive activities (brainstorming, exchanging ideas, working on drawings, conceptualizing the product, analyzing design options etc.), and it is the knowledge that flows across them and drives the decisions, shaping the product. Drawing parallels to the manufacturing environment where the raw materials get converted into the finished final product, it is the knowledge that gets converted into product design and manufacturing blue-prints in product design environment. For instance, the development lifecycle of a software product typically consists of multiple phases such as software requirement analysis, software architecture & design, coding, testing and so on. Considering the architecture phase, the high-level architecture of the software product is arrived at, detailing the various software subsystems and layers, and the key interactions and interfaces between them, taking into consideration various architectural quality attributes such as coupling and cohesion. While arriving at the architecture/design, the team analyzes the requirements and evaluates multiple options on what could be the optimal subsystems and interfaces. Based on the analysis and evaluation of the various options, decisions are made. In essence, product design and development activities can be considered to comprise a string of decisions, impelled by knowledge, taken through the various phases.

This provides additional viewpoints such as analyzing what drives these decisions, how the knowledge flows and what causes wastes.

### B. *Prevalent Knowledge & Knowledge Gaps*

A successful product can be characterized in terms of various aspects such as it being ahead of competition, on time release to the market and so on. A successful product would primarily be driven by the right and timely decisions taken during the development lifecycle of the product (e.g. decisions on the specific customer needs that are to be addressed, the right design choices to be exercised and so on). This in turn implies that the design team had the right knowledge at the right time that drove the optimal decisions. Using the term "Prevalent Knowledge" to indicate the knowledge with the design team, greater the Prevalent Knowledge is, higher is the possibility of the team making the optimal, right and timely decisions, thereby driving great products. On the other hand, poor prevalent knowledge would result in suboptimal and/or wrong decisions, causing defects, rework loopbacks among many other unfavorable consequences. Most of the problems faced during the product design and development lifecycle can be traced back to wrong or suboptimal decisions taken earlier in the lifecycle (Morgan and Liker, 2006) (Ward and Sobek, 2014) (Radeka, 2015). Sources of prevalent knowledge include individual team member, fellow colleagues, groups (when team members brainstorm in a group, they might be able to surface knowledge which might not have happened if done individually), internet/ intranet (e.g. open source forums), organizational repositories (e.g. lessons learnt databases), and external experts. There are two primary forms of knowledge: Tacit: wherein, it is in the minds of the people, and Explicit: wherein, it is in forms such as documents, drawings and pictures (Nonaka and Takeuchi, 1995). Further, product design and development involves decision making pertaining to different knowledge areas, which can be broadly classified as pertaining to functional disciplines (e.g. software engineering), technology domains (e.g. avionics), and specialty concerns (e.g. security). The knowledge available with the design team on the relevant knowledge areas for applying to the design problem at hand, constitutes the applicable prevalent knowledge for driving the decision. A gap in prevalent knowledge is said to exist if there is something that the design team does not know, but would need to know before making a decision (Radeka, 2015) (Ward and Sobek, 2014). For instance, if the design team is not sure, with a reasonable degree of certainty, that a specific software operating system would scale to meet anticipated new demands later, or if a specific communication protocol may cause performance issues in the product later, there is a knowledge gap.

### C. *Complexity, Uncertainty and Variability*

Complexity in a system is representative of the multiplex of relationships (multiplicity), diverse forces and interactions (inter-relatedness) within the various subsystems and elements, with difficulties in establishing cause-and-effect chain. Emergence, hierarchical organization and numerosity are some of the characteristics of complex systems (Ladyman et. al., 2013). Concepts of complexity across various sciences and disciplines (e.g. product design, organizational design, supply chain management) have been well researched (Jacobs and Swink, 2011). With increasing footprint of product functionality, connectivity and differentiation in new products, there is an exponential increase in the complexity associated with the design of products (Giannoccaro and Nair, 2016). Further, the product design teams are faced with significant competitive pressures pertaining to shorter design cycles and reduced costs. For complex products development, the design activities require greater depth and breadth of knowledge, across different knowledge areas. The design teams encounter difficulties in deducing the implications of the decisions on the overall product's behavior and performance. The uncertainty is due to gap in prevalent knowledge due to which the design team is typically unable to make right decisions and make real progress. Hence, complexity associated with the design of products can be summarized in terms of (a) multiplicity of the number of design decisions, (b) diversity of the knowledge areas pertaining to the decision, and (c) interrelatedness due to the significant interdependencies and multiple implications of the decisions. Figure 1 illustrates a typical scenario encountered in complex product development, wherein the knowledge to make the right decisions increases as progress is made towards the later phases of the development lifecycle. But most of the critical decisions are made during the initial phases of product development where typically the prevalent knowledge is inadequate. The progress is "illusory" when decisions are made based on opinions and wishful thinking, since invariably some of the decisions may need to be reworked later. Variability implies changes in the original factors and attributes considered during the product design. Changes midstream during product design and development may be induced due to reasons such as changing market conditions, irrelevance/ increased relevance of some of the originally envisioned product features and changes in the product perspectives from different stakeholders. Variability may require some of the decisions taken earlier to be reworked, since not changing the decision might impact product performance, competitiveness and viability in the changing market conditions. Again, for complex products, the design team would face significant uncertainty in deducing the implications of variability on the product design, and assess the changes that are required in the decisions taken earlier.

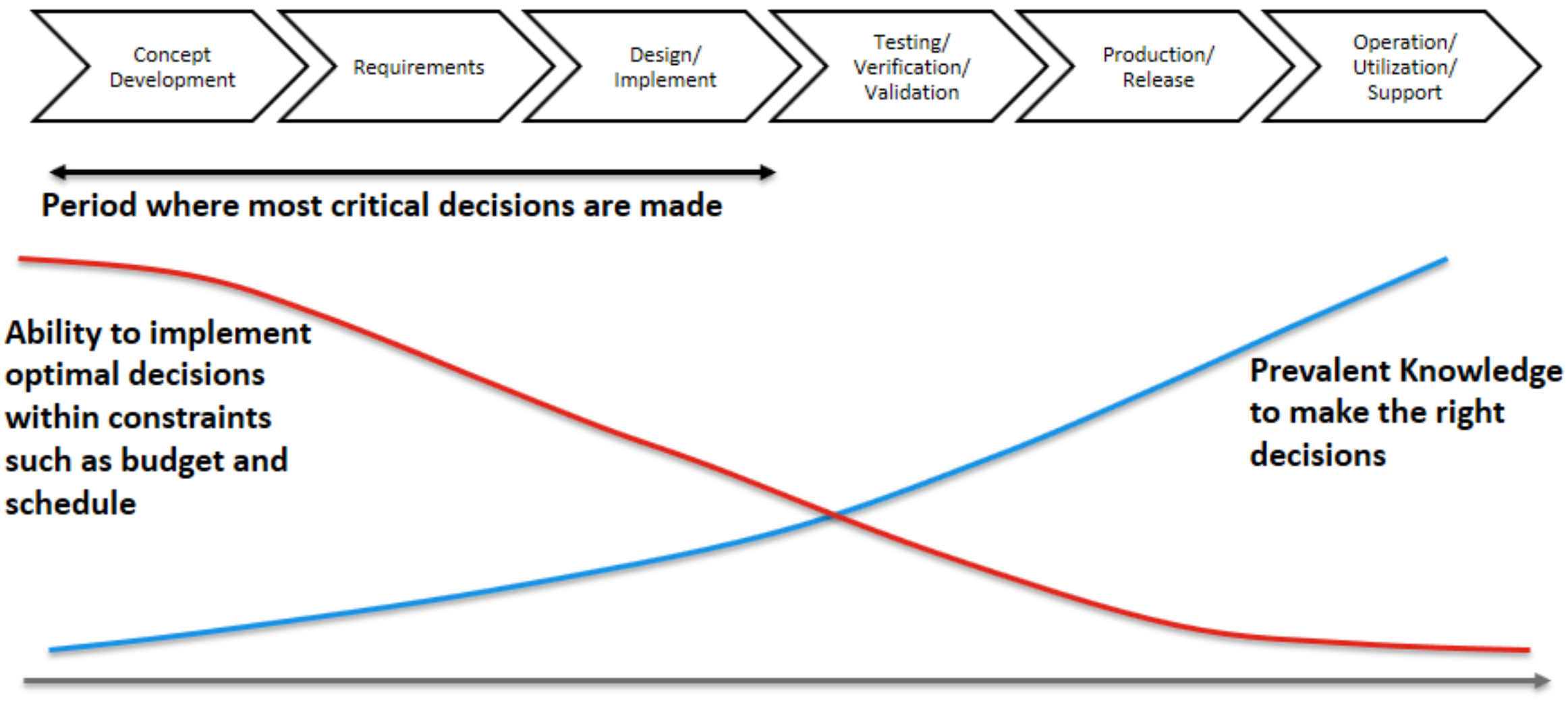

**Figure 1 Complex Product Design and Development**

### D. *Learning Cycles*

In the context of decision making in complex product design and development, the concept of Learning Cycle is of paramount importance. Learning Cycle is indicative of the consequences and duration the design team experiences before realizing the implications of the decisions taken, as illustrated in Figure 2.. As design for complex products involves decision making with significant uncertainty, the design teams encounter significant number of learning cycles. Coupled with the fact that most of the critical design decisions are made in the early development phases, when the prevalent knowledge often is not adequate to make the right decision, the ability to comprehend learning cycles becomes crucial.

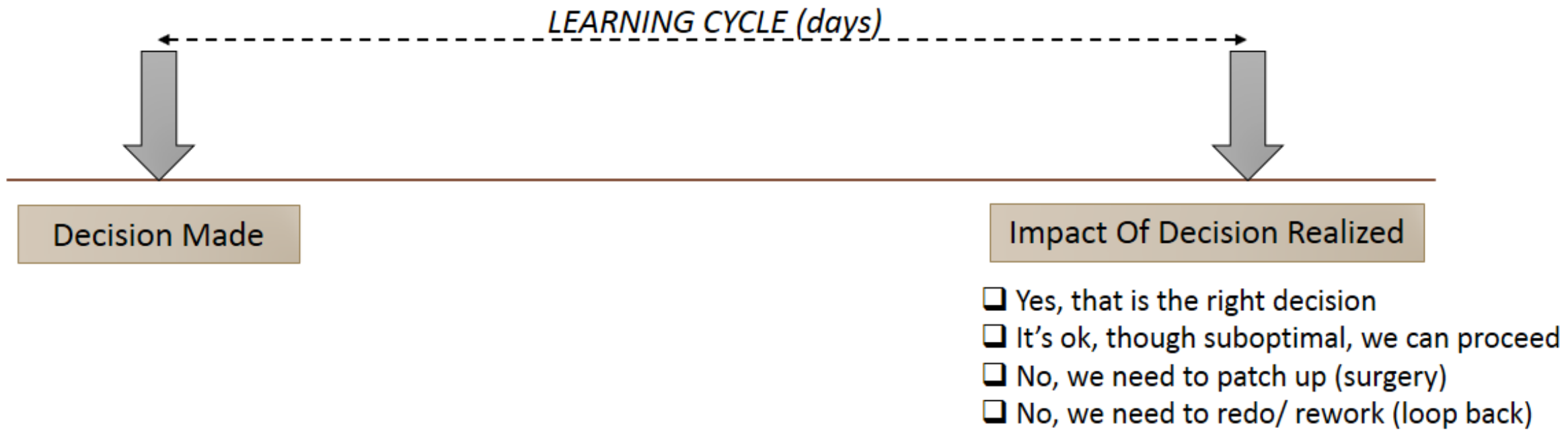

**Figure 2 Learning Cycles**

### E. *Knowledge Management in Product Design & Development*

The seminal work on knowledge aspects pertaining to product development scenario (Nonaka and Takeuchi, 1995) highlighted the importance of learning organizations, and proposed the SECI model of knowledge conversion. The SECI model of knowledge conversion comprises Socialization (from tacit to tacit knowledge), Externalization (from tacit to explicit knowledge), Combination (from explicit to explicit knowledge), and Internalization (from explicit to tacit knowledge). Subsequent extensions of SECI model have also been proposed to address various aspects and dimensions of knowledge in the context of product design and development. For instance, an integrated dynamic knowledge model was proposed (Tyagi et. al. 2015), comprising three distinct elements, viz. the SECI modes, 'ba' (Japanese word referring to space-time nexus), and knowledge assets. In product design and development, existing knowledge enhances the team's ability to understand and integrate new knowledge with existing routines and competencies (Cui et. al., 2014). Establishing ontologies for knowledge representation helps in reducing the misinterpretations associated with context connected knowledge sharing in product design (Albers et. al, 2011). Knowledge Based Engineering (KBE) systems with lean knowledge lifecycle stages including knowledge identification, representation and sharing have also been proposed. For instance, a KBE system was proposed (Sorli et. al., 2012) that allowed design departments to make decisions based on previous project knowledge and on proven technological domain knowledge, through a Lean Knowledge Lifecycle. Capturing the design space constraints was recognized as an effective means to record and evolve the knowledge associated with the product design (Ding et. al., 2011). Cognitive knowledge modeling aspects for product design pertaining to support for content, action and context of knowledge were identified in (Qiu et. al., 2008). Value from knowledge and operational value streams, waste reduction and knowledge flow/ pull are some of the principles highlighted for lean product development (Schuh et.al., 2014) (Rauch et.al., 2016).

Aspects pertaining to creation, capture, reuse of knowledge and knowledge flow in the knowledge value stream, across product development value streams, were analyzed in (Ward and Sobek, 2014), (Radeka, 2015), (Ringen and Welo, 2015), (Morgan and Liker, 2006), (Lin et. al., 2013) and (Kennedy, 2003). Usable knowledge was indicated as the basic value created in product development through three kinds of learning: integration learning (learning about customers, suppliers, usage environment), innovation learning (pertaining to new possible solutions), and feasibility learning (enabling better decisions avoiding cost and quality problems). The focus required on creating knowledge, for consistently profitable operational value streams, by Entrepreneur System Designers (project/ business line leaders) with the support of functional managers, was highlighted. Knowledge waste is highlighted as the critical waste in product development, and categorized as scatter (right knowledge doesn't get to right place, for instance due to communication barriers or poor tools), hand-off (people making decisions don't have the required knowledge, for instance, due to separation of knowledge, responsibility, action and feedback) and wishful thinking (making decisions without data or operating blindly). Tools/ processes such as LAMDA cycle (Look-Ask-Model-Discuss-Act) learning process to create knowledge, A3 reports, K-Briefs, tradeoff curves, and checksheets were recognized for knowledge based product development and knowledge based lean frameworks. For instance, A3 Thinking (Shook, 2009) reframes all problem solving activities as learning activities at every level of the organization, emphasizing on understanding causality, seeking predictability and ensuring ongoing learning. And, A3 reports serve as an appropriate means for codification of design knowledge (Raudberget and Bjursell, 2014). Value stream mapping workshops have also been recommended for eliciting tacit knowledge in product development teams (Mayrl et.al., 2013).

### *F. Addressing Uncertainty & Variability in Product Design*

There has also been significant work pertaining to uncertainty and variability in product development, in addition to frameworks and methodologies proposed for overcoming the challenges in product development. For instance, for achieving robustness in product development in face of variability, various macro and micro methods for adoption such as Axiomatic Design, Taguchi method, DFV (Design for Variety) and VRM (Variation Risk Management) are analyzed (Cabello, 2010). While overlapping upstream and downstream activities in product development is generally successful, it is difficult to predict design performances under varying overlapping policies for complex product design and development, though some insights can be gained by modeling rework (Zhang and Bhuiyan, 2015). Techniques such as Interface Management Method (Johansson and Säfsten, 2015) have been proposed to address uncertainty and complexity in product design. Uncertainty further impacts the project plans, typically requiring continuous adjustments in procurement, engineering and execution (Vaagen et. al., 2017). Studies on Toyota Product Development System (Morgan and Liker, 2006) outlined principles spanning process, people and tools/ technology dimensions. Cross functional front loading, knowledge pull through levelled process flow and design process standardization are some of the process elements. Learning and continuous improvement through kaizens, and developing towering technical competence with functional heads in the organization leading knowledge creation, ensuring knowledge capture and dissemination, are some of the people elements. Aligning the various teams in the organization through visual communication, and tools for learning and standardization, are some of the technology elements. However, many challenges in managing product development knowledge persists (Maksimovic et. al., 2014), with complexity emerging as one of the highest rated challenge across many sectors/ domains including automotive, medical equipment and software. Managing and measuring knowledge reuse, invention and exploitation is a significant challenge (Lettice et. al, 2006). Challenges pertaining to avoiding knowledge vaporization and understanding ripple effects of decisions are also discussed (Capilla et. al., 2016), while recognizing that the decision makers are prone to making errors under uncertainty (Dasanayake et. al., 2015). For instance, a comprehensive survey and analysis of architecture design decision making techniques for software products indicate that, in spite of various techniques available, decisions are often "made partly in the dark" and turn out to be suboptimal (Falessi, 2011). Decisions evolve over the product development lifecycle, and additional views such as chronological decision views are recommended (Tofan and Galster, 2014). The decision making process, associated decision theories and decision making philosophies, have also been researched (Harrison and Campbell, 2011). While architecting and designing under uncertainty and ambiguity, approaches such as set-based exploration and iterative strategy have been proposed (Gil et. al., 2008), based on whether the design team is aware of the range of architecture/ design aspects but lack the knowledge to be precise.

As can be seen from the research works discussed in sections A to F above, an integrated framework for knowledge value stream, with a deployment approach, that strings together various aspects pertaining to knowledge flow, knowledge creation and management, in complex product design and development context, specifically factoring in nuances such as uncertainty and variability, has not been adequately addressed. With trend towards increasing complexity in product design, critical aspects such as uncertainty and variability in decision making, and learning cycles need to be adequately addressed, in the context of knowledge value stream.

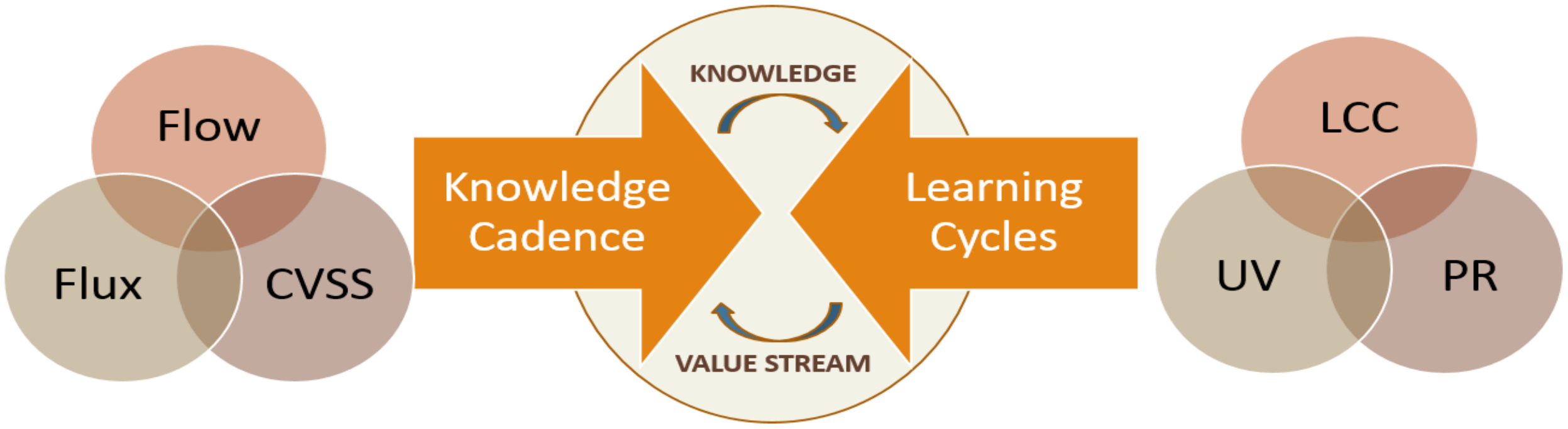

**Figure 3 Proposed Knowledge Value Stream Framework**

## III. KNOWLEDGE VALUE STREAM FRAMEWORK

This section discusses the proposed knowledge value stream framework (Figure 3) for architecting and designing complex products. Knowledge Cadence and Learning Cycles, with closed feedback loops between these two elements form the crux of the framework. Knowledge cadence incorporates Knowledge Flow, Knowledge Flux and CVSS (Create-Validate-Store-Share/Use), while Learning Cycles incorporate LCC (Learning Cycle Consequences), UV (Uncertainty-Variability) and PR (Perception-Reality) aspects.

### A. *Knowledge Value Stream Context*

Figure 4 illustrates the overall context for knowledge value stream. In the context of product design and development, the Body of Knowledge (BOK)[1] comprises all the knowledge available in the knowledge value stream, that can potentially be used by the design team to drive the design decisions. The prevalent knowledge is the relevant knowledge that is readily available to the design team towards making the decision. The performance and quality of the product can be considered as a direct reflection of the prevalent knowledge of the design team. In the case of an efficient and effective knowledge value stream, all the relevant knowledge from the BOK is available as prevalent knowledge for the design team to guide the decision. Further, knowledge value stream spans multiple product development value streams. For instance, an organization can comprise multiple Lines of Businesses (LOB) - a LOB can be considered as a set of highly related products serving a particular business need. Each of these products have a design lifecycle of their own, typically aligned with the organizational product development processes (for example, a stage-gate process). Typically, in large enterprises with multiple LOBs, the teams are organized in different structures, for instance in a functional form or in product team form. In a functional form, there could be multiple functional teams (e.g. Software Architects Team, Coding Team) and the members of those teams could move across multiple product design lifecycles. In such large enterprises, there is a wealth of knowledge spread across the wide span of the organization. Often there are huge inefficiencies in spreading, leveraging and optimizing the knowledge spread across a wide span of the organization for impactful application across product designs and developments.

[1] http://en.wikipedia.org/wiki/Body_of_knowledge

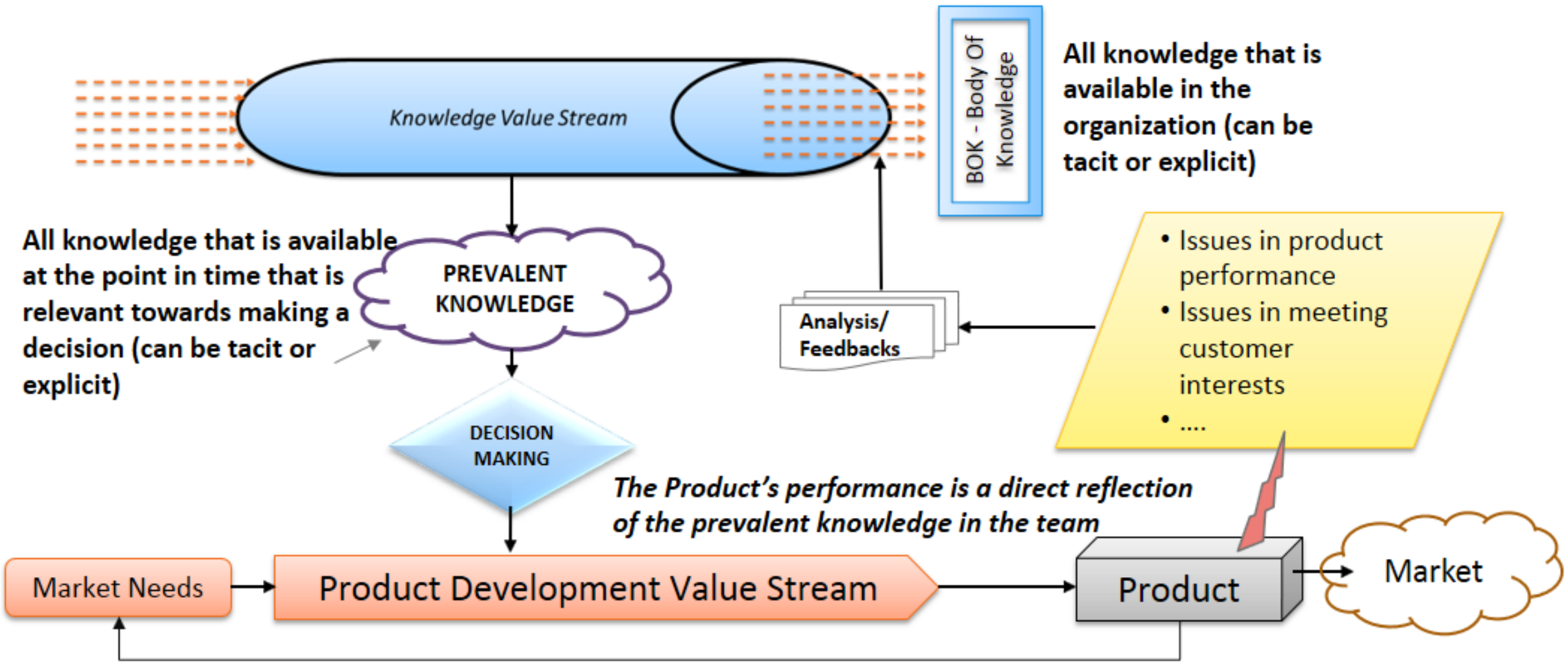

**Figure 4 Knowledge Value Stream Context**

Knowledge cadence deals with how well the knowledge flows across the multiple product development value streams, how the knowledge is created, validated, stored, shared, and how it impels the decisions during product design. Further, complex product development is characterized by significant uncertainty in decision making. This causes the design team to encounter significant learning cycles, and subsequently the consequences of such learning cycles. Learning Cycles deal with multiple scenarios of uncertainty and variability occurring during product design and development, and varying perceptions of what real knowledge gaps are, and consequently the different consequences faced.

### B. *Knowledge Flow & Knowledge Flux*

Figure 5 illustrates a scenario of how knowledge flows in a design team. In this example, there are 4 members in the team. A directed edge (knowledge tie) from John to Raj indicates that John approaches Raj towards getting knowledge for making decisions. The nodes are referred to as knowledge actors. When knowledge actors are people, they represent sources of tacit knowledge. Else, they represent sources of explicit knowledge (for instance, "database" in the figure is a source of explicit knowledge). This approach for modeling knowledge flow is further discussed in subsequent sections. This paper also proposes the concept of knowledge flux in the knowledge value stream, thereby providing the interrelation between knowledge flow and decisions. In the simplest terms, knowledge flux can be viewed as similar to magnetic flux in the physical world (flow of energy through a surface). In other words, knowledge flux represents the extent of knowledge flow that impels the decisions being taken in the product development value stream. Figure 5 illustrates how the knowledge flux can be assessed. The knowledge flux indicated in the figure includes all the knowledge ties, and hence represents the total knowledge flux. When only a subset of the knowledge ties pertaining to tacit knowledge flow is considered, the knowledge flux pertaining to tacit flow can be assessed. The form of knowledge in the knowledge value stream has a bearing on the optimal levels of knowledge flow and knowledge flux measures. If the proportion of tacit knowledge is high (say, greater than 80%), there needs to be strong people to people networks. If the learning cycle consequences are also favorable in this scenario (i.e. mostly, the decisions taken were found to be optimal for the product), this would imply that team members work very well closely, with very good interactions, and make optimal design decisions. On the other hand, if the proportion of explicit knowledge is high, and if the learning cycle consequences are favorable, it would imply that most of the members of the design team often rely on accessing the explicit knowledge, say a decision guide database, to drive the design decisions.  Figure 6 illustrates the tacit – explicit knowledge form implications on design decisions.

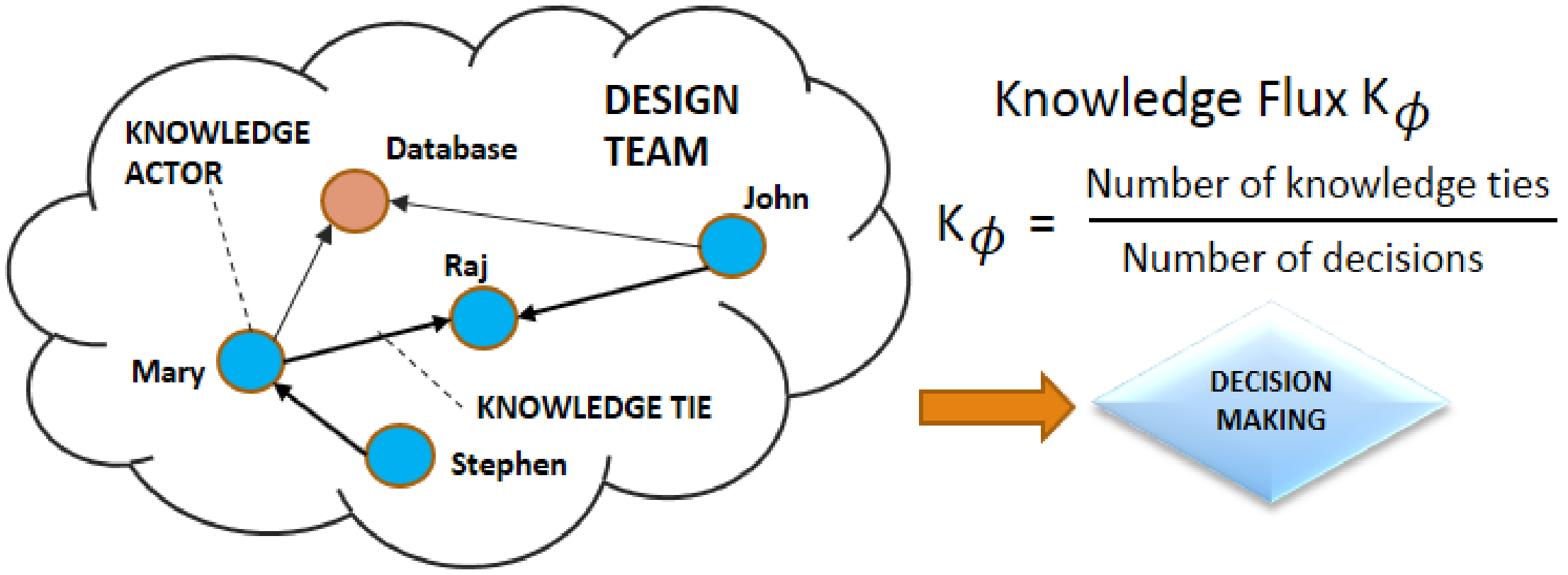

**Figure 5 Knowledge Flow & Flux**

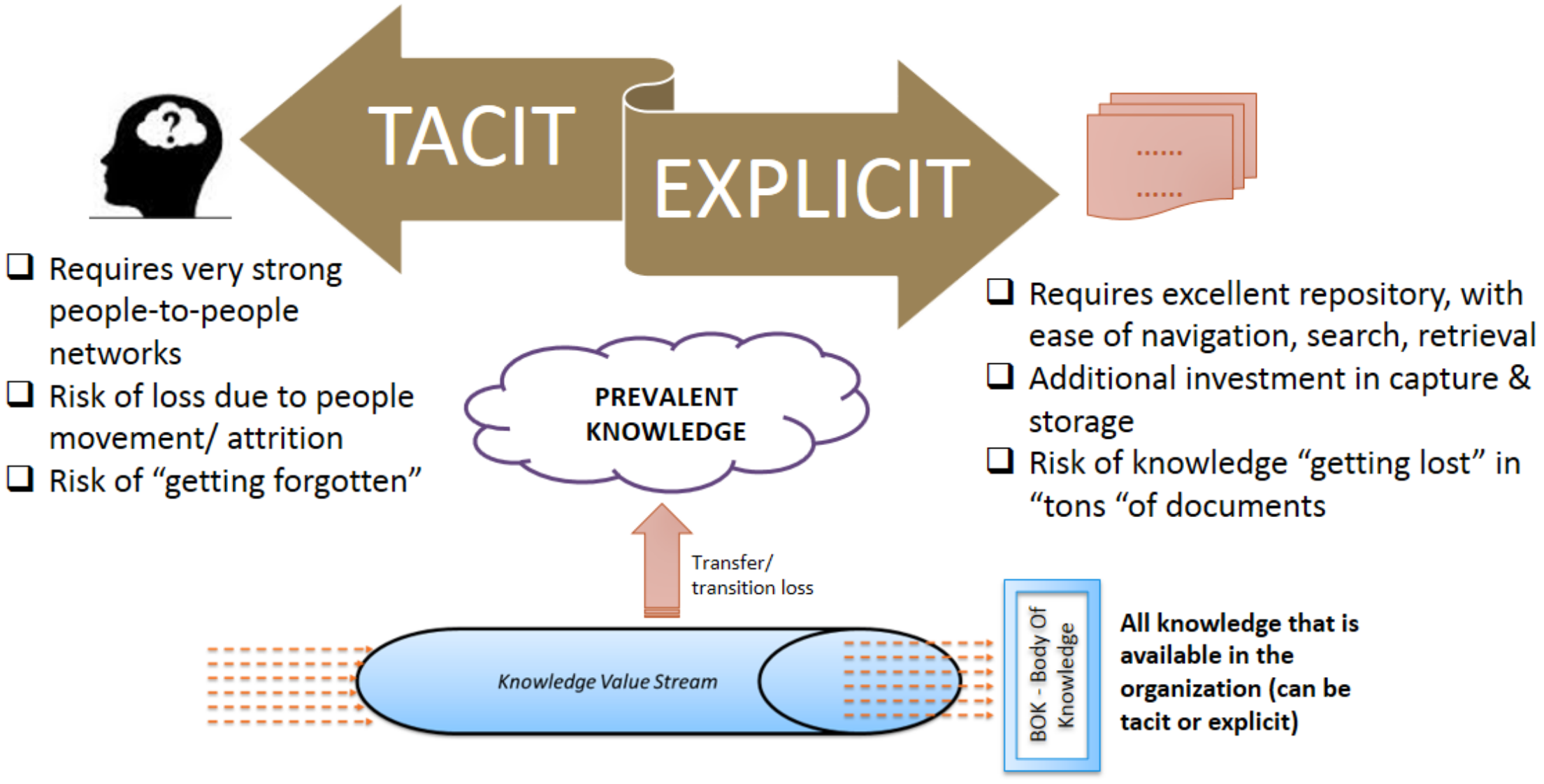

**Figure 6 Implications of Tacit and Explicit Forms of Knowledge on Knowledge Flow/ Flux**

*C. Create-Validate-Store-Share/Use Elements*

The different elements in knowledge value stream pertaining to creation, validation, storage, and sharing and usage of prevalent knowledge, are illustrated in Figure 7. New knowledge created, as a result of closing knowledge gaps and as a result of learning cycles, constitutes the creation element. Validation occurs when learning cycles are encountered on the decisions taken. Storage refers to the means where the knowledge is stored for retrieval, depending on whether tacit or explicit. For instance, in an organization, most of the knowledge might be in an explicit form, in databases. Sharing refers to how the knowledge gets shared, for instance through knowledge sharing sessions. Usage refers to how the knowledge base gets used by the design teams. Wishful thinking is said to occur when the decisions are made without closing the knowledge gaps, or when the teams are under the perception that they have the required knowledge. This often results in "illusory" progress being projected. For instance, if a decision on using a specific communication protocol is made without adequate leverage of knowledge pertaining to the protocol available in BOK and/or not addressing the knowledge gaps pertaining to that decision, it is highly likely that unfavorable learning cycle consequences (discussed later) are encountered. Hence, the development progress assumed earlier due to that decision being taken, gets invalidated.

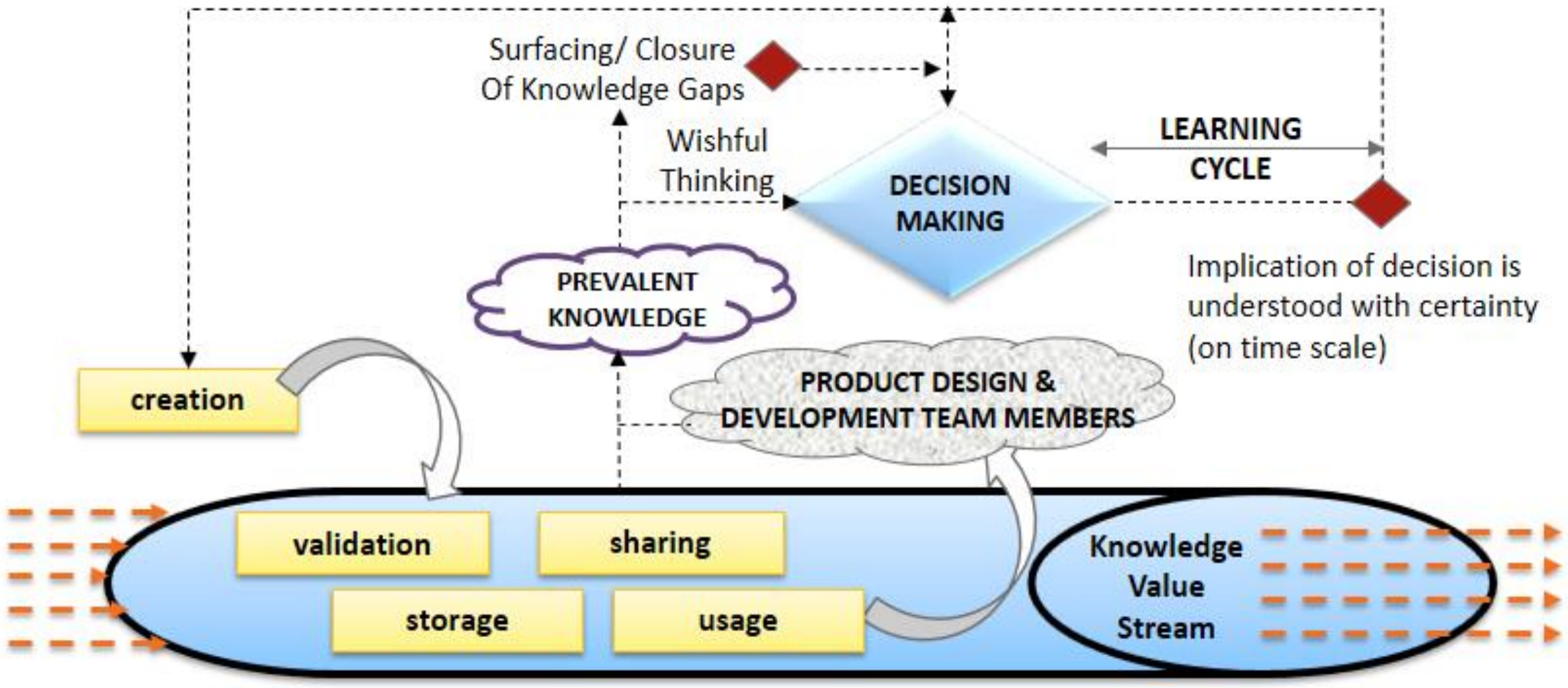

**Figure 7 CVSS – Create, Validate, Store, Share/Use Elements**

*D. Uncertainty & Variability*

As discussed earlier, in product design and development, uncertainty is due to gap in prevalent knowledge due to which the design team is unable to make right decisions and make real progress (flow) in product development. There are three typical scenarios that occur in product development, pertaining to uncertainty and variability, as illustrated in Figure 8.

(a) UV-1: The first scenario wherein the customer, market, business needs are well understood, is indicated by a star in needs space. No variability is expected in the needs. Prevalent knowledge is sufficient to clearly identify and design the optimal solution (solution space). Hence, there is very less uncertainty, as indicated by a star in the solution space. Typical product development approaches (e.g. waterfall, point based serial engineering, point based concurrent engineering), would be most appropriate. (b) UV-2: In the second scenario, uncertainty is significant in the solution space, indicated by a cloud i.e. the prevalent knowledge is not sufficient to clearly identify & design the optimal product. Traditional product development approaches, if followed, might indicate good progress initially, but during later part of the lifecycle, significant effort might be consumed on making the chosen design work, and this would often result in a sub-optimal solution. Product development approaches based on working on sets of solutions, and gradually converging on strongest solution by elimination (e.g. adopting Set based Design (Ström, 2016)) might be more appropriate. Closure of knowledge gaps (predominantly technical) becomes crucial. (c) UV-3: In this scenario, uncertainty and variability is significant, as indicated by clouds in needs and solution space.  The prevalent knowledge is not sufficient to narrow down the customer needs and to identify and develop the optimal solution (product). Product development approaches based on faster iterations and gaining customer insights (e.g. Lean-Startup (Ries, 2011)), with flexibility and configurability in the solution would be appropriate. Rapid Learning Cycles (Radeka, 2015) and learning milestones leading to closure of knowledge gaps (pertaining to technical, customer, market etc.) become more relevant.

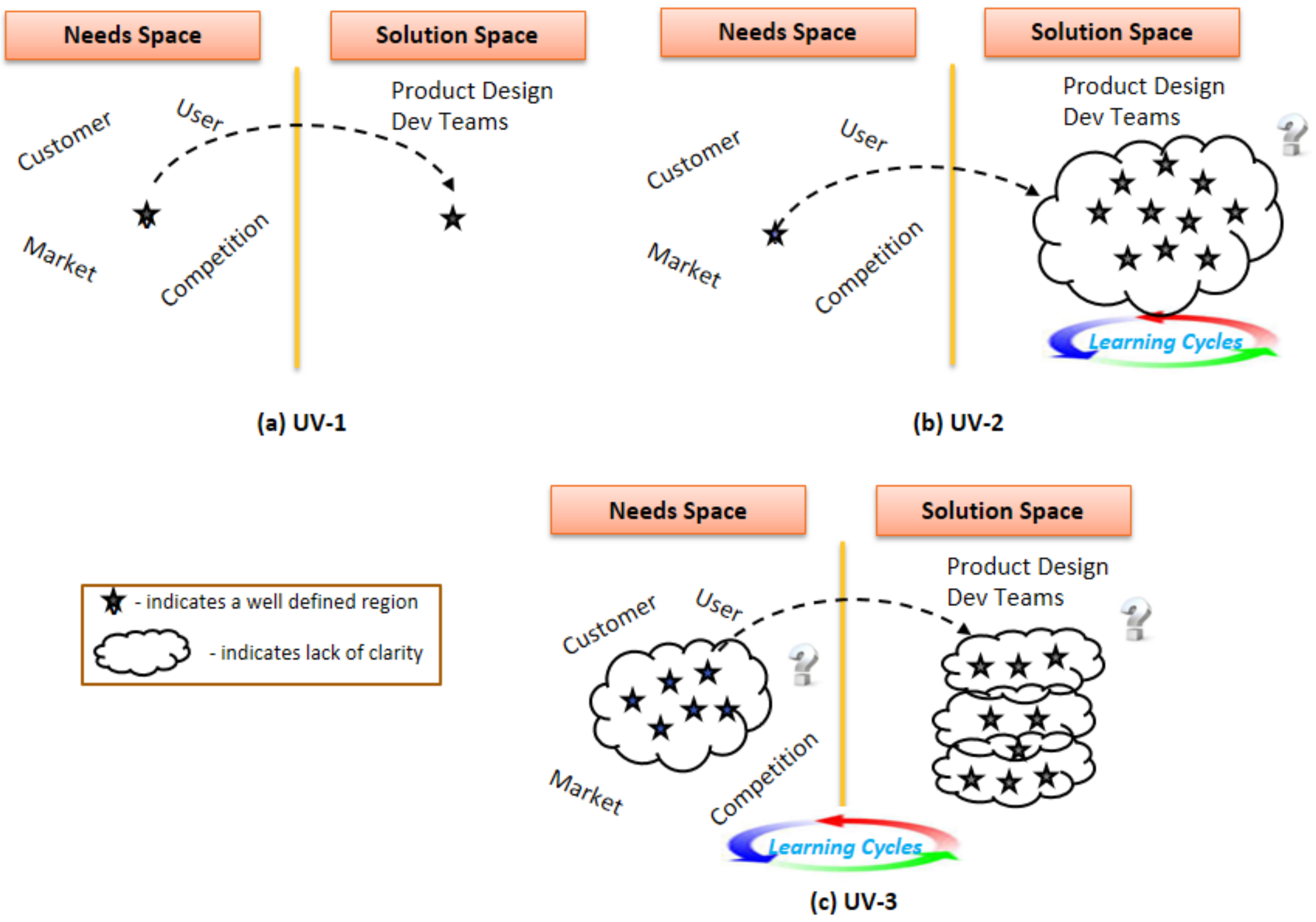

**Figure 8 Uncertainty – Variability Scenarios**

## E. *Learning Cycle Consequences*

Learning Cycle (Figure 9) is indicative of the duration the  design team experiences before realizing the implications of the decisions taken. When the team makes design decisions based on the knowledge available at that point in time when the decision has to be made, the Learning Cycle comes into play by impacting further development and subsequent product performance. For instance, when the team realizes that an architectural decision needs to be reworked, when performance problems surface during integration testing, the team encounters a learning cycle consequence. Table 1 enlists the different learning cycle consequences. The duration of the Learning Cycle also matters - shorter cycles are always better than longer ones. While assessing the uncertainty for a specific decision to be taken during product design, the learning cycle and feedback loops are to be taken into account. For instance, longer learning cycles with LC-4 consequence implies higher uncertainty, as illustrated in Figure 9.

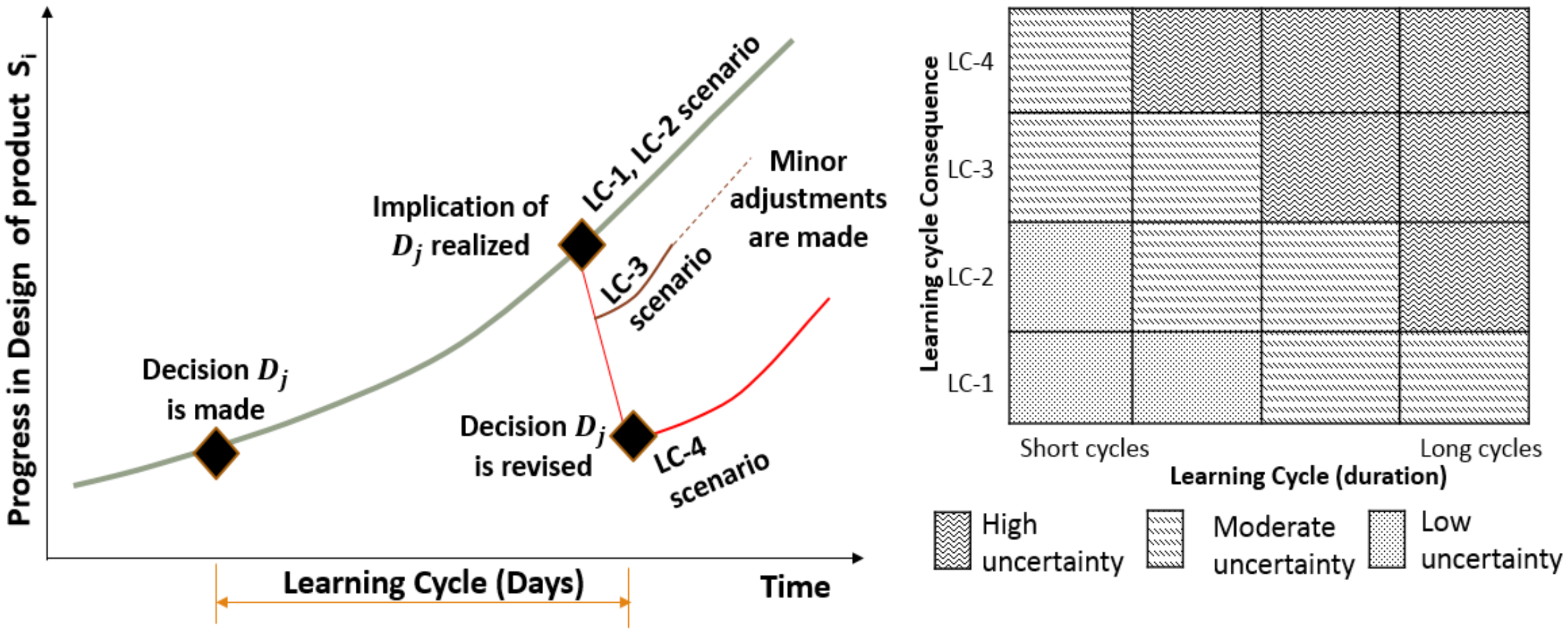

**Figure 9 Learning Cycles**

**TABLE 1: LEARNING CYCLE CONSEQUENCES**

| | |
|---|---|
| LC-1 | The decision is the optimal decision, and does not inhibit any of the requirements or expected behaviors of the product |
| LC-2 | The decision is not the optimal one, but nevertheless, can be “lived with”, i.e. does not impact any critical requirements or behaviors of the product |
| LC-3 | The decision is not optimal, and it might require some amount of rework / minor correction/ “surgery” |
| LC-4 | The decision needs to be significantly reworked, requiring a loop-back to the point where the decision was taken. In extreme cases, the budget/ resources required for the rework might be far in exceedance of what is available/ allowable |

## F. *Perception-Reality*

Perception implies the way in which something is regarded and understood. Reality implies the state of things as they actually exist. Often, an inefficient and ineffective knowledge value stream will give rise to perceptions that differ from reality, for the design team. The impact of perception can be in two categories.

Illusory Progress: Examples of perceptions of the design team that induce illusory progress are: “we know the real customer interests well...”, “we know the right design choices to be made for this product…”, “we know exactly the design specifications and how the integration is to be done…” and “we know this will not cause any issues downstream…”. However, the reality might be that the prevalent knowledge with the design team is inadequate (i.e. there are knowledge gaps) to make the right decisions. The design team presumes that the decisions made are optimal, and proceeds ahead. This perception impels the team to make the suboptimal/ wrong decisions, thereby causing unfavorable learning cycles and rework later in the development lifecycle.

Hence, the progress assumed early (when the decisions were made) is termed "illusory". Figure 10 illustrates this scenario where the teams are under an illusion that they know sufficiently about the needs space (problem) and the solution that is required (design decisions) for the product. For instance, it might be the scenario that another team in the organization faced some issues while attempting a solution on a similar problem. An inefficient knowledge value stream might have caused loss of that knowledge, or might have caused that knowledge not reaching the said team at the required time. In such a scenario, the design teams are not aware of the knowledge gaps, and are under the perception that they are making the right optimal decisions. However, later in the development lifecycle, the teams realize that those were not the right decision. In such scenarios, significant wastes manifest during the product development by situations such as addressing the wrong customer interests and/or making the wrong design decisions, projecting an illusory progress, and subsequently designing the wrong solution.

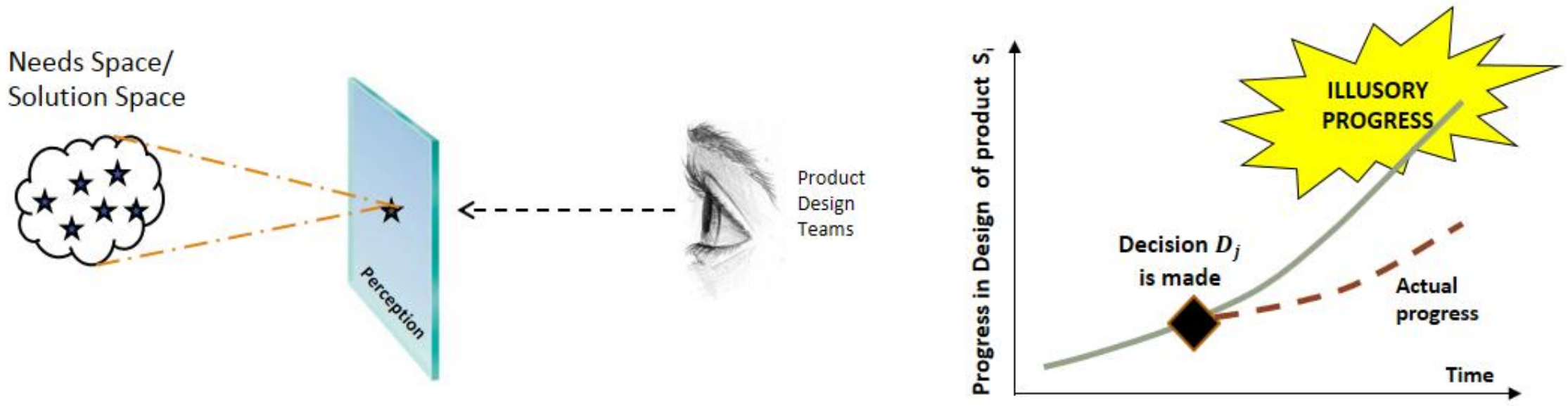

**Figure 10 Perceptions - Illusory Progress**

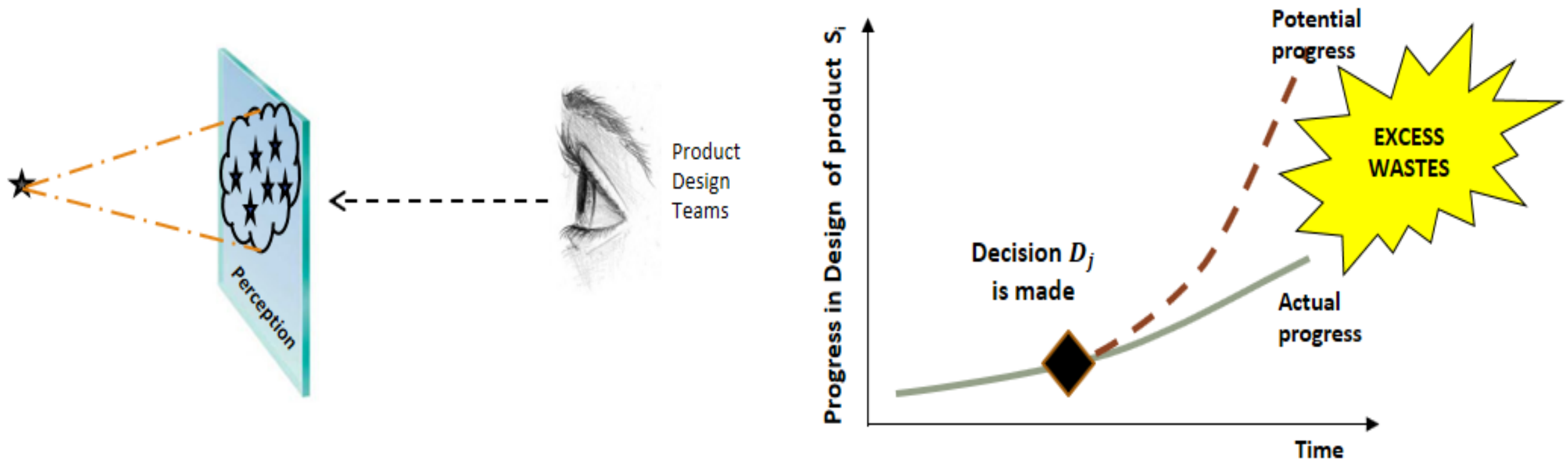

**Figure 11 Perceptions - Excess Wastes**

Excess Wastes: The second category of impacts due to perception is Excess Wastes. Examples of perceptions that induce excess wastes are: "we are not sure of the real customer interests well...", "we are not sure if the design choices are optimal…", "we are not sure of how the integration is to be done smoothly…", "this might cause some issues downstream…let's wait…". These cause the design team to spend more effort and resources on overcoming the perceived knowledge gaps – for instance, by exploring multiple solutions. For instance, it might be the scenario that another team in the organization had faced a similar problem, and had already solved the problem successfully. An inefficient knowledge value stream might have caused loss of that knowledge, or the required knowledge not reaching the design team at the right time, thereby causing this design team to "re-invent" the solution. Figure 11 illustrates the scenario of perceptions causing excess wastes thereby slowing down the progress in the development lifecycle, as against the potential progress if the knowledge value stream was efficient and effective. The impact of perceptions hence can essentially be viewed from the perspective of knowledge gaps. Figure 12 illustrates this perspective, depicting scenarios of perception causing the prevalent knowledge to be perceived more or less than the actual prevalent knowledge.

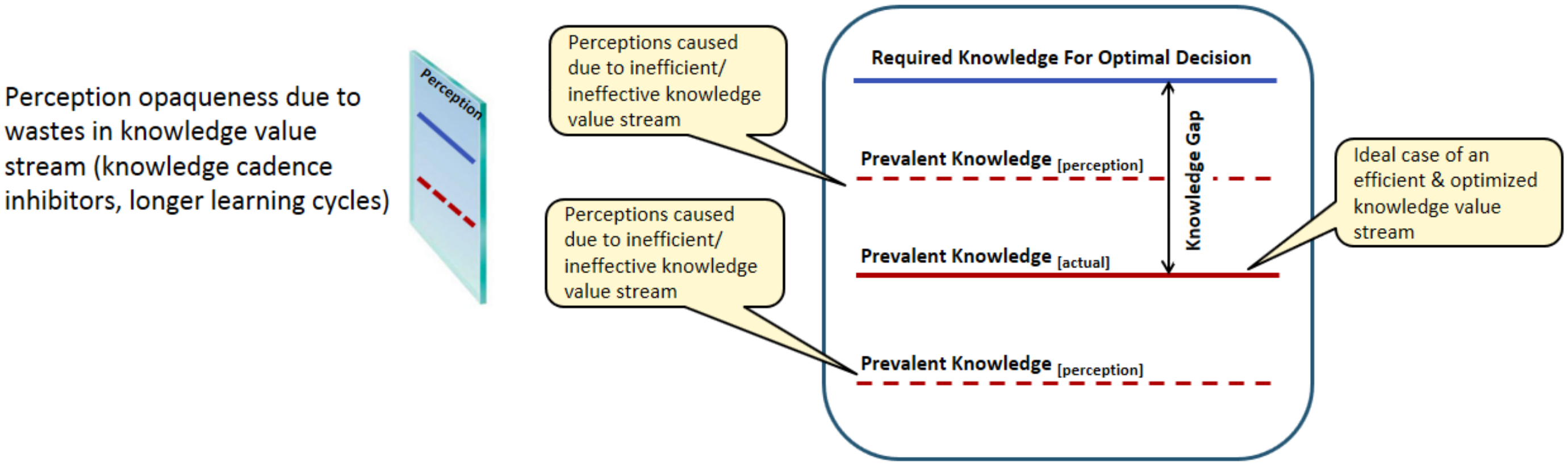

**Figure 12 Perceptions - Knowledge Gaps**

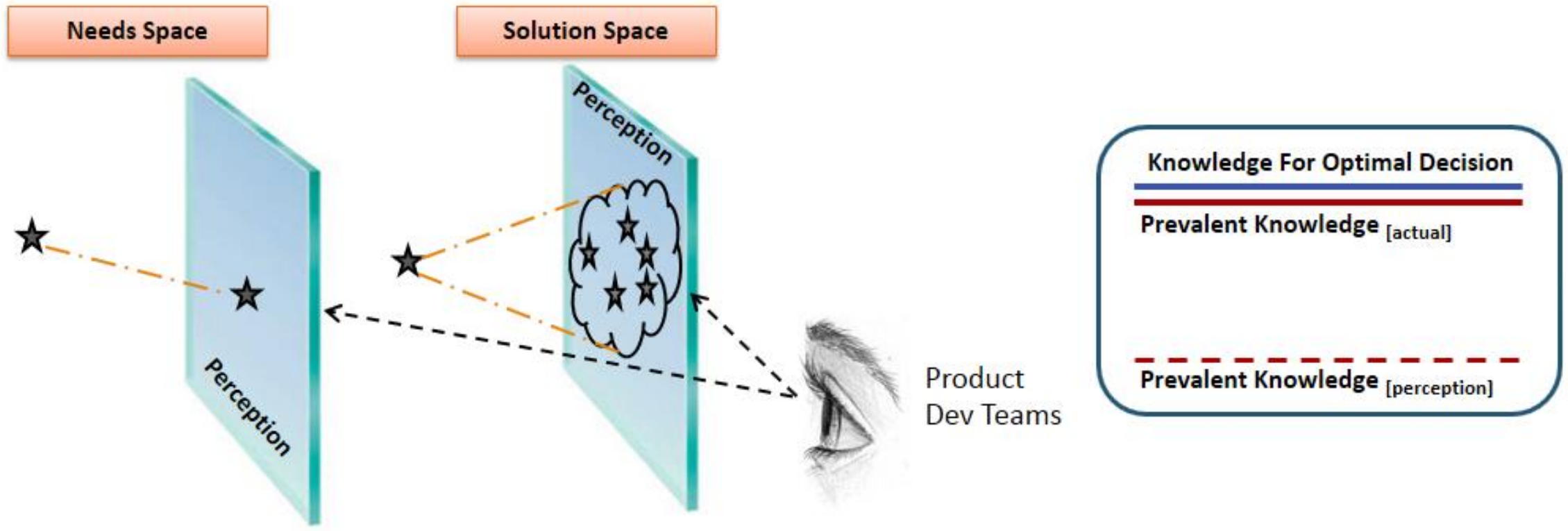

**Figure 13 Perception-Reality Scenario**

The "opaqueness" in the perception (causing a difference with the reality) is caused due to wastes in the knowledge value stream, such as wastes causing the required knowledge not reaching the team at the required time. There might be scenarios wherein the perception differs from reality in only one of the spaces (needs space or solutions space). For instance, Figure 13 illustrates the case where the perception differs from reality in the solution space only. This scenario is wherein the design team has a very good understanding of the needs space, but are under the perception that their prevalent knowledge is inadequate to arrive at the optimal solution. This induces moderate wastes in the product development value stream, due to the team exploring more to finalize the optimal design. This might impact the cycle time of the product development and/or the time to market.

### *G. Closed Loop: Knowledge Cadence – Learning Cycle*

The degree of uncertainty associated with a design decision might gradually decrease as the learning on the implications of the decision increases. The learning increases as the implications are understood better, through one or more of the following scenarios: (a) when more knowledge is gained through prototypes, and simulations of the product, components or any specific portion of the product, (b) when subsequent development phases do not encounter any problem that are traceable to the specific decision, and (c) when actual test, field demonstration, and deployment do not encounter any problems that are traceable to the specific decision. Establishing the closed loop in the knowledge value stream is critical to ensure that the BOK in the organization matures and evolves over a period of time. When the implications of a design decision taken are realized, the BOK needs to incorporate the learnings due to that decision, as new knowledge. Specifically in large organizations, different teams learn similar lessons over different points in time – which indicates that there is a loss of knowledge. Establishing the closed loop ensures that the knowledge is fed back to be taken cognizance of in future decisions, and that there is no knowledge loss. Post the deployment/ release, the product's performance is a reflection of the prevalent knowledge the team had in driving the decisions in the needs space and solution space. Hence, a closed loop reflection on the product's performance in terms of the original decisions that were taken, again generates new knowledge and/or validates existing knowledge. This ensures that the BOK matures progressively over a period of time.

### *H. Framework Elements: Inter-relationship View*

Figure 14 illustrates the overall view of inter-relationships between the key elements of the framework. The formalism associated with these key elements are discussed subsequently, to establish the required foundation for deployment approaches based on the proposed framework. Let $P^U$ denote the set of all products that were designed and developed, or those currently under development in the organization. Let $K^U$ represent the set of all knowledge areas relevant for designing the various products in the organization. Let A be the set of all knowledge actors in the organization (Knowledge Actors can be people or databases/ repositories etc.). $Z^+$ :set of positive integers.

$$P^U = \{P_1, P_2, .., P_n\}, \quad n \in Z^+ \ldots\ldots\ldots\ldots\ldots\ldots\ldots\ldots\ldots\ldots\ldots (Def.1)$$

$$K^U = \{K_1, K_2, .., K_m\}, \quad m \in Z^+ \ldots\ldots\ldots\ldots\ldots\ldots\ldots\ldots\ldots\ldots\ldots (Def.2)$$

$$A^U = \{A_1, A_2, .., A_q\}, \quad q \in Z^+ \ldots\ldots\ldots\ldots\ldots\ldots\ldots\ldots\ldots\ldots\ldots (Def.3)$$

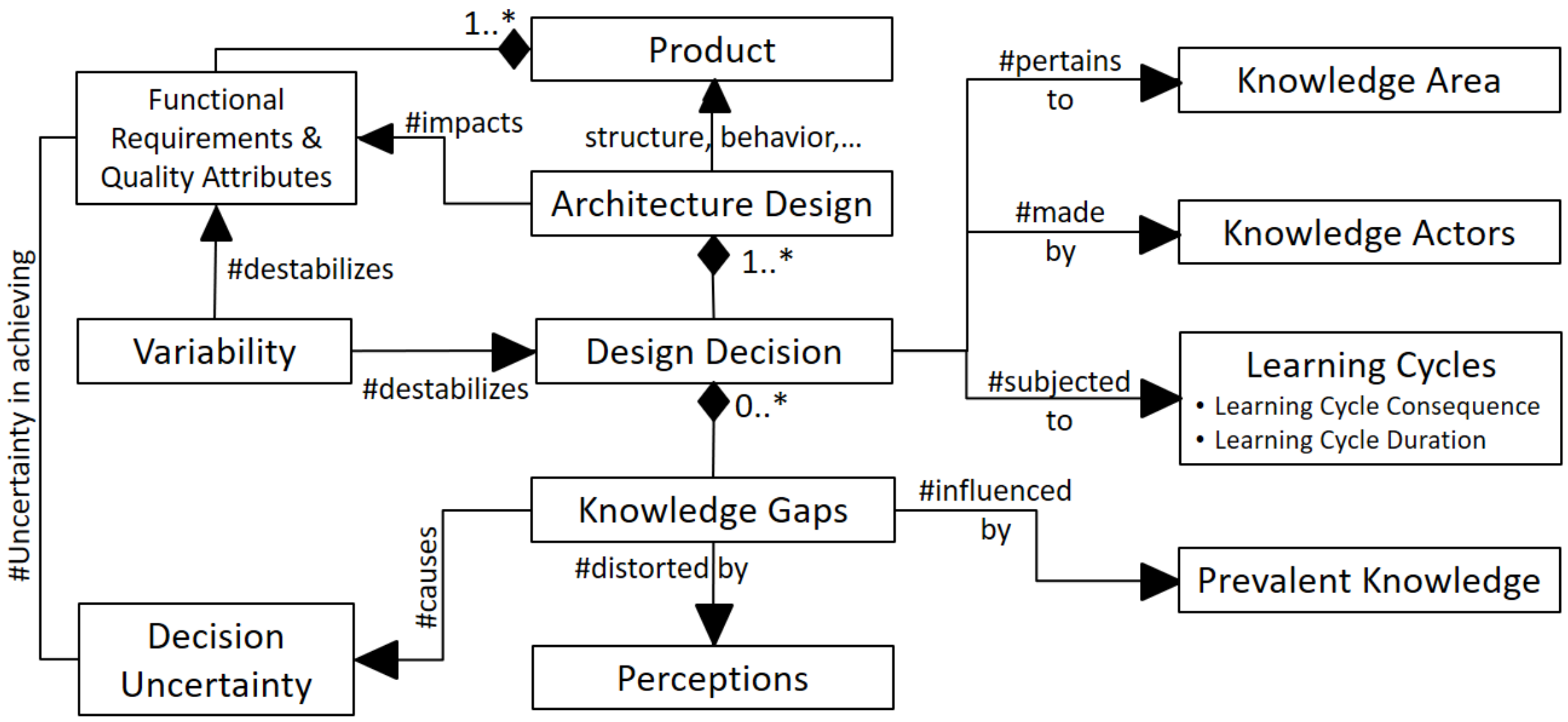

**Figure 14 Framework Elements: Inter-relationships View**

Designing a product $P_j$ involves making a set of design decisions $D^{Pj}$ , which pertain to different knowledge areas. A subset of knowledge actors are involved in making the decision.

$$D^{Pj} = \{ d_1^{Pj}, \ldots d_v^{Pj} \}, \ldots v \in Z^+ \ldots\ldots\ldots\ldots\ldots\ldots\ldots\ldots\ldots\ldots\ldots\ldots (Def. 4)$$

For complex product design, Learning Cycles are experienced for each of the decisions, in terms of one of the learning cycle consequences (LC-1 through LC-4) and duration (e.g. short, medium, long). The below function captures the learning cycle experiences associated for decision made for product Pj

$$LCC = \{(LCC_1, duration), \ldots (LCC_4, duration)\} \ldots\ldots\ldots\ldots (Def.\ 5)$$

$$Function\ fLCC^{Pi} :\ D^{Pj} \ \rightarrow LCC \ \ldots\ldots\ldots\ldots\ldots\ldots\ldots\ldots\ldots\ldots\ldots (Def. 6)$$

During the design of complex products, there are significant knowledge gaps associated with decisions. Let $G^{d_v^{Pj}}$ be the set of knowledge gaps associated with making decision $d_v^{Pj}$. A set of knowledge actors $A^{d_v^{Pj}} \subseteq A^U$ are involved in making the decision.

$$G^{d_v^{Pj}} = \left\{ g_1^{d_v^{Pj}}, g_2^{d_v^{Pj}}, \ldots\ g_r^{d_v^{Pj}} \right\}, r \in Z^+, d_v^{Pj} \in D^{Pj} \ldots\ldots\ldots\ldots (Def. 7)$$

The perceived set of knowledge gaps $G_\lambda^{d_v^{Pj}}$ by the actors for making the decision may be more or less than the actual knowledge gaps, causing a difference between the perception and reality. When there is no difference between perception and reality, it pertains to the ideal case of an efficient and effective knowledge value stream. However, there are various scenarios that might occur, based on the relationships between the perceived set of knowledge gaps and the actual knowledge gaps. For instance, if the perceived set of knowledge gaps is a proper subset of the actual set of knowledge gaps, the scenario depicted in Figure 10 will be encountered, that might result in illusory progress. On the other hand, if the actual set of knowledge gaps is a proper subset of the perceived set of knowledge gaps, the scenario depicted in Figure 11 will be encountered, that might result in excess wastes.

Efficient knowledge value stream: $G_\lambda^{d_v^{Pj}} = G^{d_v^{Pj}}$

A scenario of Illusory progress: $G_\lambda^{d_v^{Pj}} \subset G^{d_v^{Pj}}$

A scenario of excess wastes: $G^{d_v^{Pj}} \subset G_\lambda^{d_v^{Pj}}$

To facilitate different uncertainty representations/ models, the proposed framework considers the universal set of uncertainty levels, $UT^U$ as a partially ordered set, with $u_i \preccurlyeq u_j$ implying that uncertainty level of $u_i$ is less than $u_j$.

$$UT^U = \{u_1, .. u_t\} \quad t \in Z^+, (\ UT^U, \preccurlyeq\ )\ is\ a\ poset \ldots\ldots\ldots\ldots\ldots\ldots (Def.\ 8)$$

$$Function\ fUT: D^{Pj} \rightarrow UT^{Pj}, UT^{Pj} \subseteq UT^U, \left|UT^{Pj}\right| > 0 \ldots\ldots\ldots\ldots (Def.\ 9)$$

There can be multiple options for incorporating the uncertainty in the decision model. For instance, one option is to use a graded qualitative scale, like “Low Uncertainty” and “High Uncertainty”, another is to represent a percentage scale assessment of the uncertainty, which could be based on the number/ nature of knowledge gaps involved in the specific alternative. Interval, convex sets and imprecise probability are some of the uncertainty representations/ theories discussed in literature (Nikolaidis et. al. , 2011).

## IV. FRAMEWORK DEPLOYMENT APPROACH

This section discusses some elements of the deployment approach for the proposed framework outlined in the previous section. Sections A to D discuss approaches for deploying framework elements pertaining to Knowledge Cadence (Knowledge Flow, Knowledge Flux and CVSS). Sections E to F illustrate approaches for deploying framework elements pertaining to Learning Cycles (Learning Cycle Consequences, Uncertainty-Variability and Perception-Reality). Approaches discussed include modeling and analyzing the knowledge flow and knowledge flux, and assessing the processes associated with creation-validation-storage-sharing of knowledge. Further, an illustration of how a phase-wise deployment can be structured is also discussed, in addition to progressively maturing the prevalent knowledge, and ensuring that the knowledge value stream aligns with required innovation.

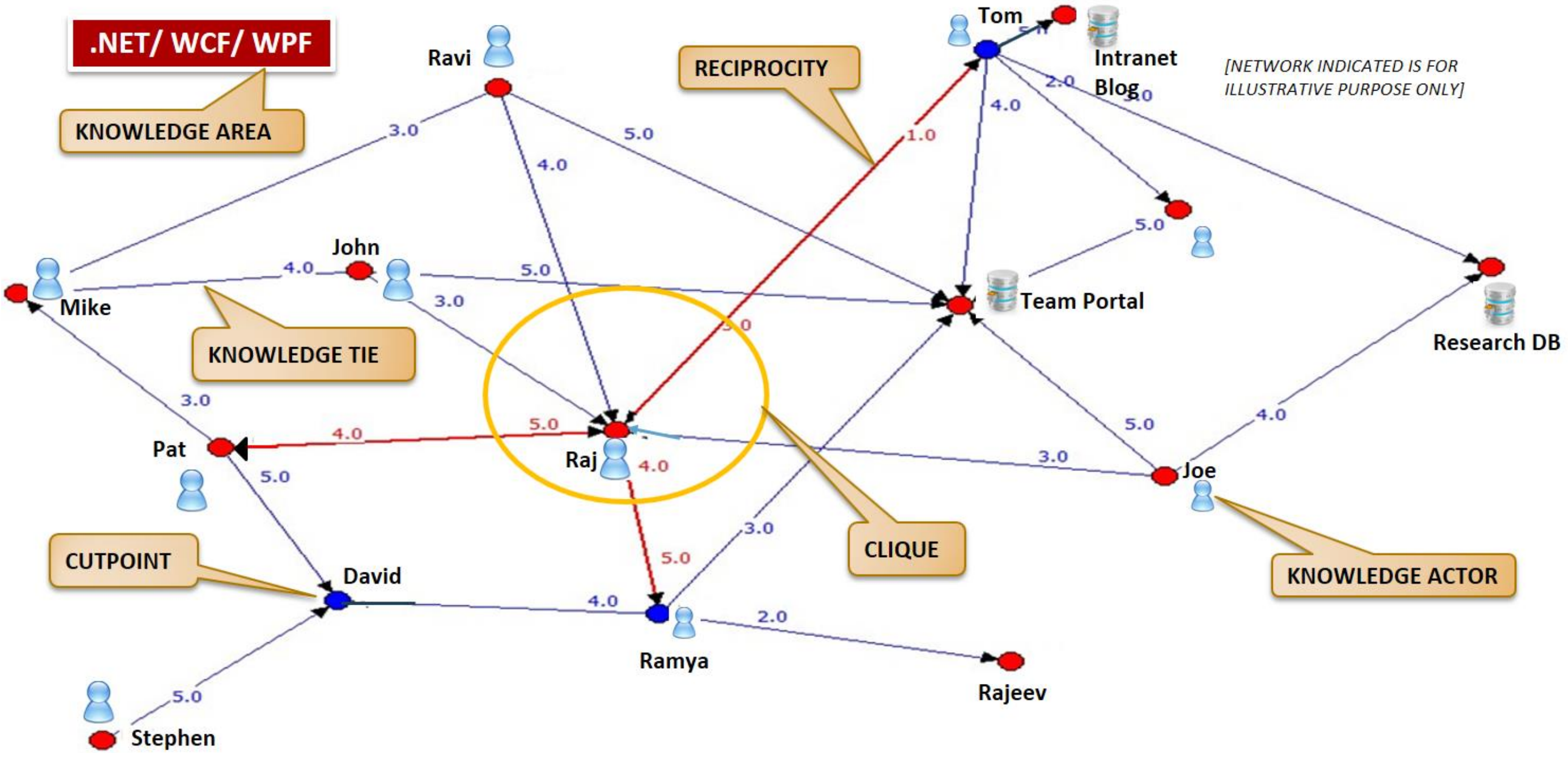

**Figure 15 - Modeling Knowledge Flow**

### A. *Modeling Knowledge Flow*

Modeling knowledge flow that is prevalent in a team provides a view of how knowledge flows amongst the various sources of knowledge to drive decisions. One of the means to model and analyze knowledge flow for a team, adapted from (Chandra et. al., 2015), is illustrated in Figure 15. This modeling can be done for each knowledge area – the example illustrated in the figure indicates the knowledge area as ".NET/WPF/WCF" (as discussed earlier, a knowledge area represents a domain of knowledge which is of relevance for the design of a product). The nodes indicate a knowledge source (for instance, an individual or a repository) termed as a "Knowledge Actor". In the figure, directed edges from John to other knowledge actor nodes indicate that when John has a problem or question or needs more information on ".NET/WPF/WCF" decision, during the design of the product, he approaches the "Team Portal", Mike, and Raj. These edges are referred to as "Knowledge

Ties". The Knowledge ties can be assigned weights to indicate preferences on who is approached first - for instance the knowledge tie from John to "Team Portal" has a high number 5 (indicating a higher preference), as opposed to knowledge ties from John to other knowledge actors. When a person approaches another person, it indicates a flow of tacit knowledge. When a person approaches a repository, it indicates a flow of explicit knowledge. The resulting graph can be analyzed based on different perspectives. For instance, in the example illustrated, if David left the organization, it would result in Stephen getting disconnected from other team members, since Stephen approaches only David mostly. In this case, David can be considered as a "Cut Point".

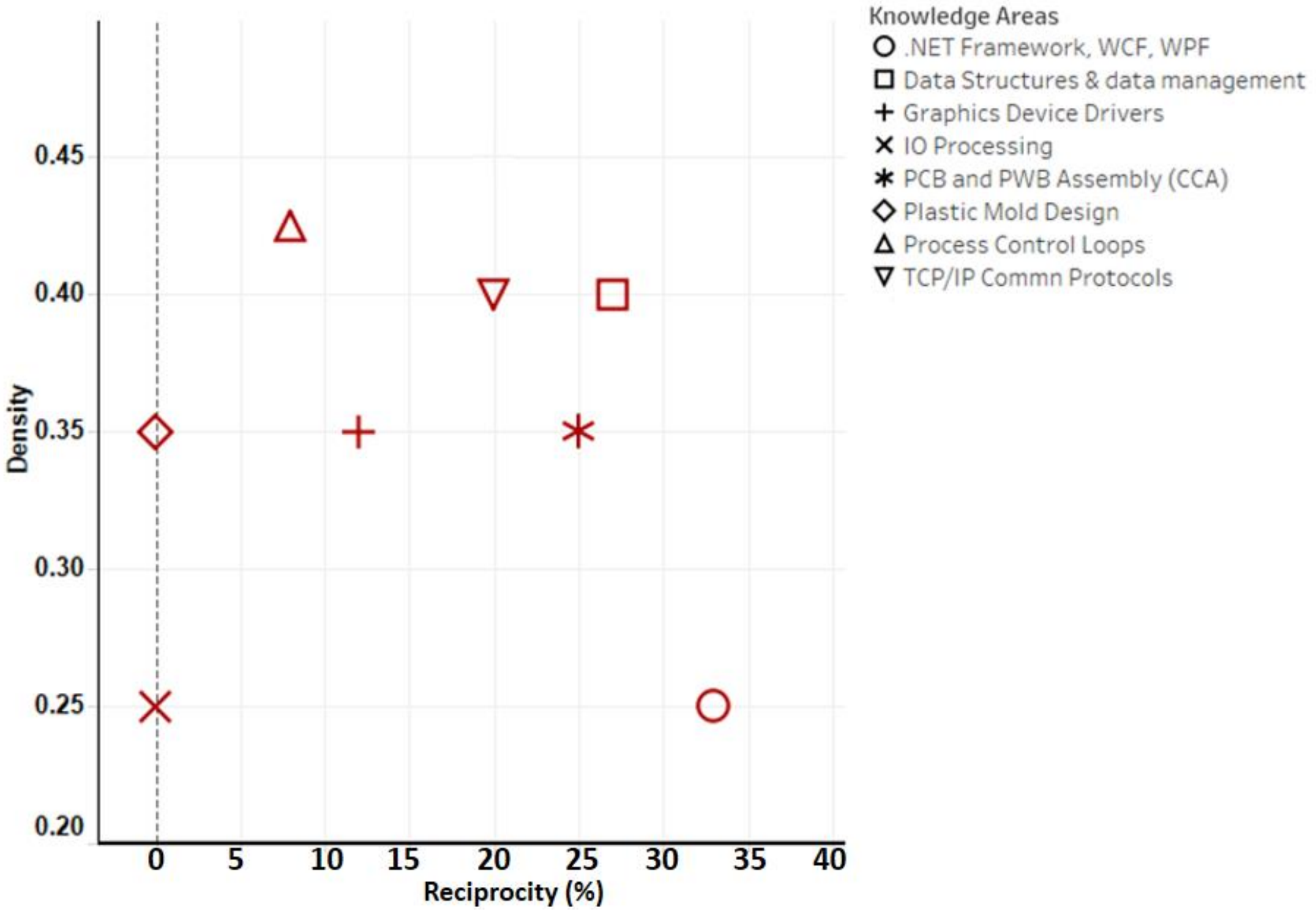

**Figure 16 Density - Reciprocity Analysis**

Bi-directional edges indicate that the knowledge actors approach each other, and this can be assessed as a measure indicating reciprocity. Having modeled the flow in this manner, one can assess how well the knowledge flows in this team, for each knowledge area. Bidirectional edges among a set of three or more nodes indicates a clique. It is ideal if there is a clique between the experts of that knowledge area – which provides an apt case for forming a "Community-Of-Practice" for that knowledge area. Various measures pertaining to the knowledge flow can be assessed. The emerging patterns of knowledge flow can thus be analyzed through measures such as density, reciprocity, number of "cutpoints", and tacit versus explicit. For instance, high value of density and reciprocity measures among people knowledge actors, and higher value of knowledge flux, would imply strong people to people networks. TABLE 2 illustrates the example measures. There are other possible ways too, to analyze the measures. For instance, a comparison of the density and reciprocity can be done to analyze the presence of bi-directional interactions (Figure 16), which can facilitate formation of groups and communities of practice. Knowledge areas with low reciprocity but with good density can be considered as areas where focused effort to improve the bidirectional interactions can provide quicker benefits. By analyzing various measures and emerging flow patterns for different knowledge areas, specific actions may be identified to ensure that the knowledge flow is efficient with minimal waste, and to mature knowledge flow on technical design capabilities.

**TABLE 2: EXAMPLE MEASURES FOR ANANLYZING KNOWLEDGE FLOW**

| | |
|---|---|
| Density | Actual connections of possible person-person connections (value ranges from 0 to 1) |
| Reciprocity | Bidirectional people connections (i.e. A to B, and B to A) of the actual connections (value ranges from 0% to 100%) |
| Most approached | Person/ resources approached by significant number of people |
| Cut Points | Persons, who if left the team, cause a disconnect in the flow |
| Tacit vs Explicit | People to people edges vs People to non-people edges (value ranges from 0 to 100%) |

### B. *Knowledge Flux*

As discussed earlier, knowledge flux represents the extent of knowledge flow that impels the decisions being taken in the product development value stream. Knowledge flux can be assessed by analyzing the number of knowledge ties driving the decision. It is to be assessed per knowledge area. For instance, while modeling the knowledge flow for knowledge area pertaining to serial communication protocols, if the number of knowledge ties turn out to be 25, and if the number of decisions pertaining to serial communication protocols during the design of a specific product is 100, the knowledge flux would be 0.25 (i.e. 25/100). The optimal knowledge flux for a specific knowledge area would result in most of the decisions having learning cycle consequences LC1 or LC2. If, for a team, most of the decisions taken pertaining to a specific knowledge area, are encountering learning cycle consequences LC3 or LC4, actions may be needed to enhance the knowledge flux, including the level of prevalent knowledge. This might involve bringing in the required knowledge to close the knowledge gaps, for instance through external experts, thereby increasing the number of knowledge ties (and consequently, the knowledge flux). As discussed earlier, the form of knowledge (tacit or explicit) also has a bearing on the optimal knowledge flow/ flux measures. For instance, low flux measures might be optimal if proportion of explicit knowledge is high, and the teams often encounter favorable learning cycles. Further, there might be scenarios where a team transitions, over a period of time, from a high tacit knowledge proportion to a high explicit knowledge proportion, thereby resulting in a corresponding transition curve of the optimal values for flow and flux measures. Figure 17 illustrates an example of how knowledge flux can be analyzed. The plot depicts the knowledge flux for a set of knowledge areas pertaining to the design of a product, for instance, plastic mold design being done by a team of mechanical engineers. Say, the knowledge flow analysis indicates 69 knowledge ties among 18 knowledge actors. If the product involves 75 decisions pertaining to plastic mold design, it would imply 0.92 (i.e. 69/75) as the knowledge flux assessment.

The question arises if this is the optimal knowledge flux or not. In this case, the knowledge flux is not optimal, since only about 22% of the time, the decisions made by the team pertaining to plastic mold design, have been found to be optimal, or something that does not mandate a change. Specific actions may need to be taken to enhance knowledge flux, in terms of enhancing the number of knowledge ties, bringing in additional knowledge actors (e.g. experts, additional explicit knowledge sources such as design databases) into the decision-making network. On the other hand, decisions pertaining to "IO Processing" knowledge area have predominantly (>70%) encountered favorable learning cycle consequences, and hence the knowledge flux level for that can be considered as optimal. The figure illustrates the recommended direction of movement towards the ideal knowledge flux for the specific knowledge areas.

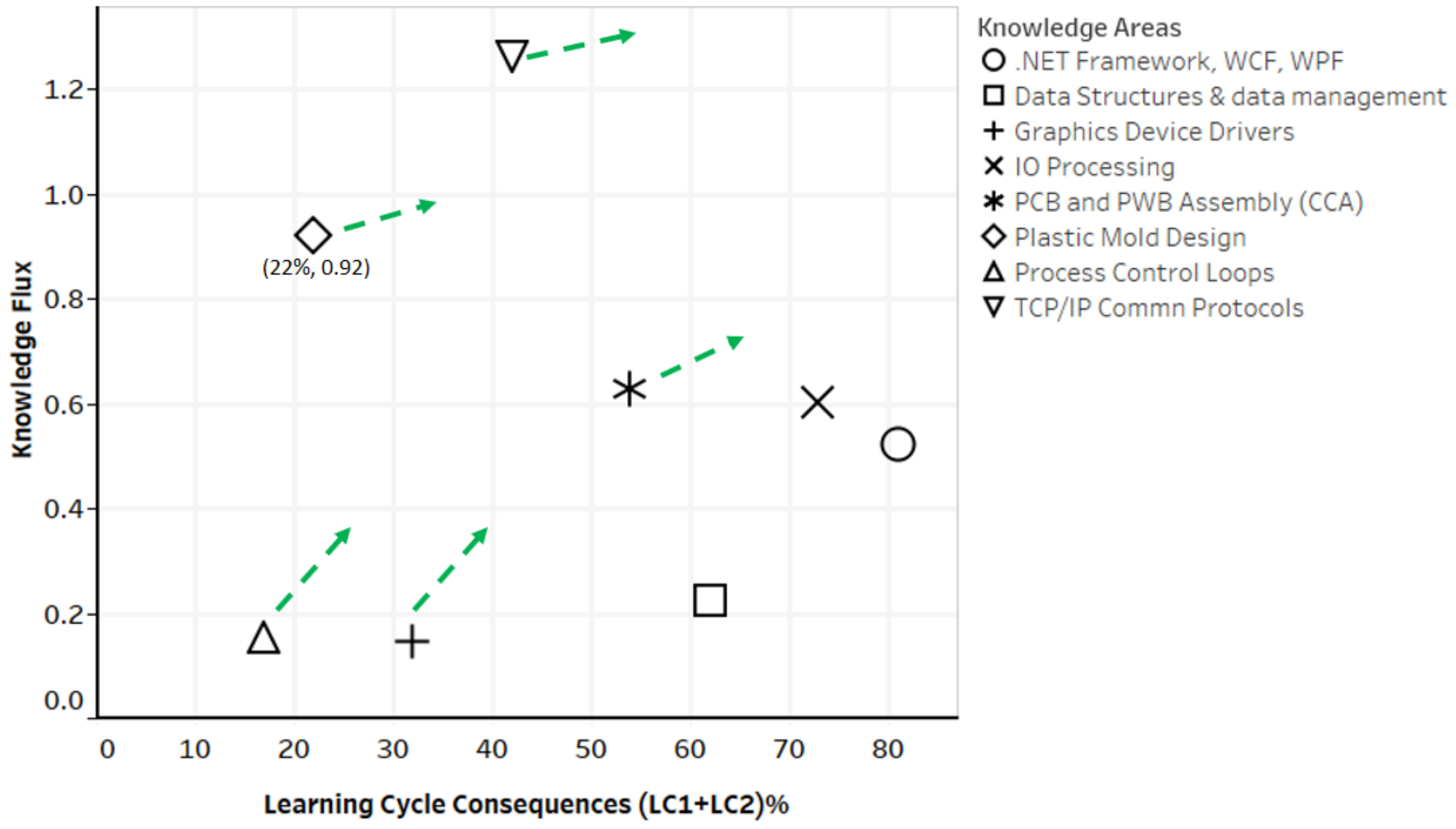

**Figure 17 Knowledge Flux Analysis**

## C. *Knowledge Flow – Knowledge Flux Analysis Reports*

An assessment of the knowledge flow and knowledge flux may be done periodically for the various product design teams in an organization across applicable knowledge areas. Based on the modeling and analysis results, specific inferences can be made, and actions identified to improve the maturity of the knowledge value stream. Figure 18 illustrates a sample Knowledge Flow - Flux analysis report for a set of knowledge areas.

| KNOWLEDGE AREA | DENSITY | RECIPROCITY | TACIT:: EXPLICIT | KNOWLEDGE FLUX | KEY OBSERVATIONS FROM NETWORK GRAPH | HEALTH ASSESSMENT |
|---|---|---|---|---|---|---|
| FPGA Design | 0.28 | 14% | 72%:28% | 0.24 | Encourage more reciprocity or mutual knowledge sharing within the team; Need to Increase reliance on explicit Body of knowledge( K- briefs/ A3s) in order to increase explicit knowledge percentage; | RED |
| Graphics Device Drivers | 0.45 | 31% | 12%:88% | 0.54 | No single point failures. Interactions between experts is good. Need to ensure awareness in team on decision databases. Evolution of knowledge to be directed. | YELLOW |
| Data structures | 0.81 | 60% | 64%: 36% | 0.72 | Presence of cliques - Conditions ideal for initiating community of practice. Attrition risk of loss of knowledge with presence of cut-points | GREEN |
| COM/ DCOM | 0.32 | 15% | 32%:68% | 0.27 | Single Point Failure Risk - Actions needed to spread their knowledge to others and also make their knowledge more explicit | YELLOW |

**Figure 18 Example of Knowledge Flow – Flux Analysis Report**

## D. *CVSS Maturity Assessments*

CVSS aspects can be matured by formulating a maturity assessment model for the teams. Table 3 indicates a maturity assessment scorecard for illustrative purposes. The scores are at four levels – Strongly Agree (SA), Agree (A), Disagree (D), Strongly Disagree (SD), with corresponding scores of 9,7,3 and 1. The score for a dimension is arrived at based on the rating given by the team for each aspect. For instance, the score for CREATE will be ((SD + D + A + SA) – 4)/32, which is 50%. The score ranges can be mapped to a scale: <25% as Weak; 25% to 50% Marginal; 51% to 80% Effective; and, > 80% as Robust. Similar score cards can be done for other dimensions. Such assessments can be done periodically, with a target score being set for the team, to gradually improve the maturity. This would enable systematic actions to be arrived at to enhance the maturity of each dimension, and subsequently assess the impact of the actions in achieving the required maturity. The duration taken for improvement would vary, depending on the maturity of the team on the CVSS.

### TABLE 3 EXAMPLE CVSS SCORECARD

| **Attribute – CREATE** | **Score** |
|---|---|
| All team members/stakeholders actively contribute to usable Body of Knowledge, and metrics are in place to assess their contributions | SD |
| Limits of existing technical design/process capabilities are established and monitored | D |
| Work standards/ Operating Procedures/ Guidelines/ Instructions are created, maintained and made available to the team | A |
| Critical knowledge assets, that reduce dependency on experts and dependency on members moving out of the team, are created | SA |
| **Attribute – VALIDATE** | |
| Review process/ workflow is in place for validation of knowledge assets created (e.g. training materials, Product & Process guidelines) | SD |
| Experts validate all critical knowledge assets | A |
| Knowledge assets are periodically revalidated (retire obsolete assets, upgrade assets etc. ) | A |
| Knowledge assets are validated with external benchmarks and gaps assesed | SA |
| **Attribute – STORE** | |
| Repository is in place for storing knowledge assets | SA |
| Repository enables easy classification of assets | SA |
| Repository has easy navigation/ searching of assets | A |
| It is common practice to ensure any new knowledge asset is stored in the repository | A |
| **Attribute – SHARE** | |
| List of experts on critical knowledge areas are published, and are easily approachable | A |
| Knowledge sharing sessions (formal/ informal) are held periodically | D |
| Teams leverage collaboration tools, actively seek inputs from experts, capture learnings in knowledge sharing sessions | A |
| Incentives are in place to motivate sharing of knowledge. Knowledge hoarding is discouraged | A |
| **Attribute – USE** | |
| It is common practice to check knowledge base before enlisting knowledge gaps or making decisions | D |
| Decisions are predominantly driven from the knowledge base – and if the knowledge is not present in the knowledge base, and work is done to close the knowledge gap, it is common practice to capture that knowledge back into knowledge base | D |
| Measures/ metrics are in place to assess extent of usage of knowledge assets | SD |
| Leaders promote knowledge based decisions, and mentor decision making using knowledge base | A |

## *E. Analyzing Learning Cycle Consequences*

Designing a complex product involves decisions spanning multiple knowledge areas. Each decision can be represented as a set of attributes with associated values. Hence, the decision space can be considered as a multi-dimensional space, with each dimension representing an attribute. A decision can thus be represented as values of a set of attributes (quantitative or qualitative). For instance, a decision "Bend Finish" pertaining to "Plastic Mold Design" knowledge area, can be represented by various attributes such as "Bending Radius", "Injection Rate" and so on with each attribute having values – for instance, "Bending Radius = 45 degrees". Another example is for the attribute "Metal-type" pertaining to another decision, which can be set with values such as 1 to 4, with metal-type 1 representing a specific aluminum alloy, metal-type 2 representing a specific steel alloy and so on.

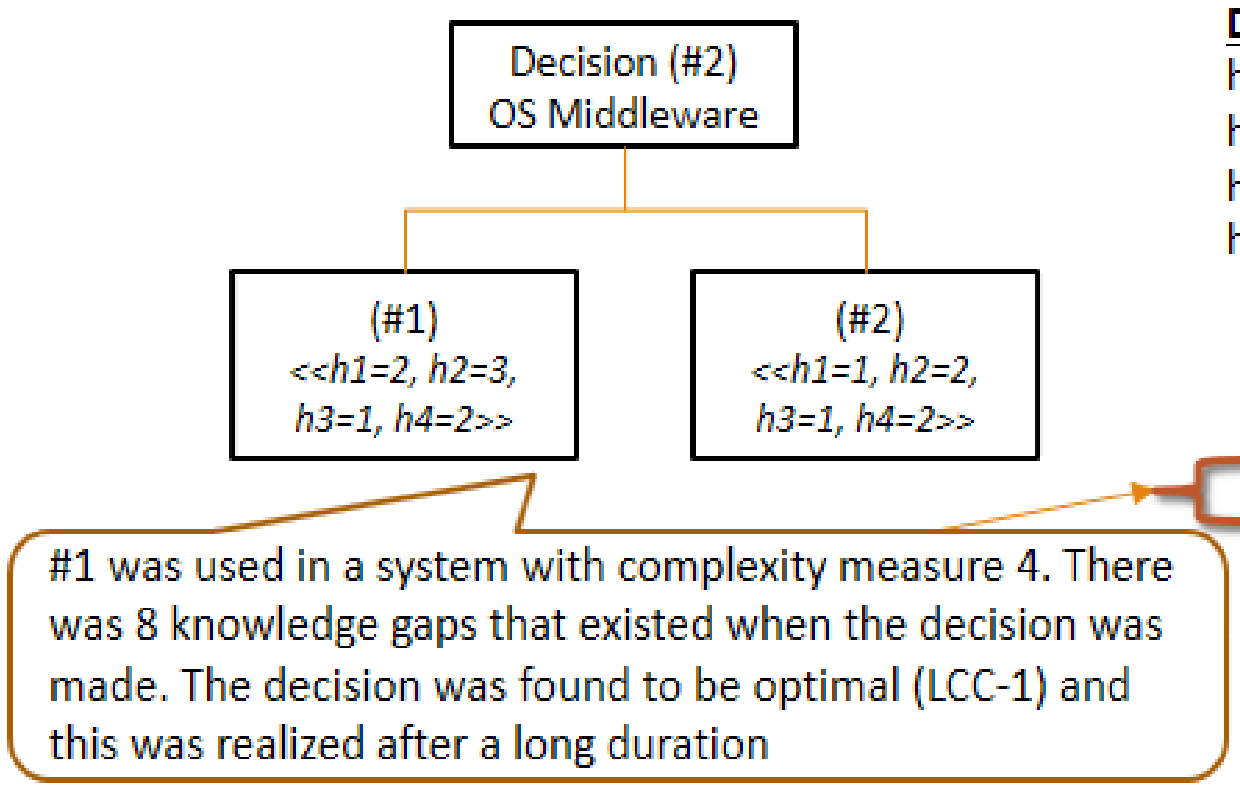

**Decision #2: OS Middleware : - Attributes**

h1 Scheduling: co-operative = 1, rate monotonic=2,...

h2: NVM writing: periodic = 2. on demand = 3,...

h3: Buffer: fixed = 1, variable=2, ring buffer=3,...

h4: thread life management: ....

Codification of Learning Cycle Experience pertaining to Design decisions

| DecisionID | Knowledge Area | h1 | h2 | h3 | h4 | Complexity | K-Gaps | LCC | LCD |
|---|---|---|---|---|---|---|---|---|---|
| 2 | OS Middleware | 2 | 3 | 1 | 2 | 4 | 8 | 1 | 3 |
| 1 | TCP Stack | 1 | 2 | 2 | 1 | 3 | 12 | 3 | 2 |
| 3 | Plastic Mold | 2 | 3 | 1 | 2 | 1 | 5 | 3 | 2 |
| 1 | TCP Stack | 1 | 2 | 2 | 1 | 2 | 6 | 2 | 1 |
| 2 | OS Middleware | 1 | 2 | 1 | 2 | 3 | 11 | 4 | 3 |

**Figure 19 Codification of Design Decisions - Learning Cycle Consequences**

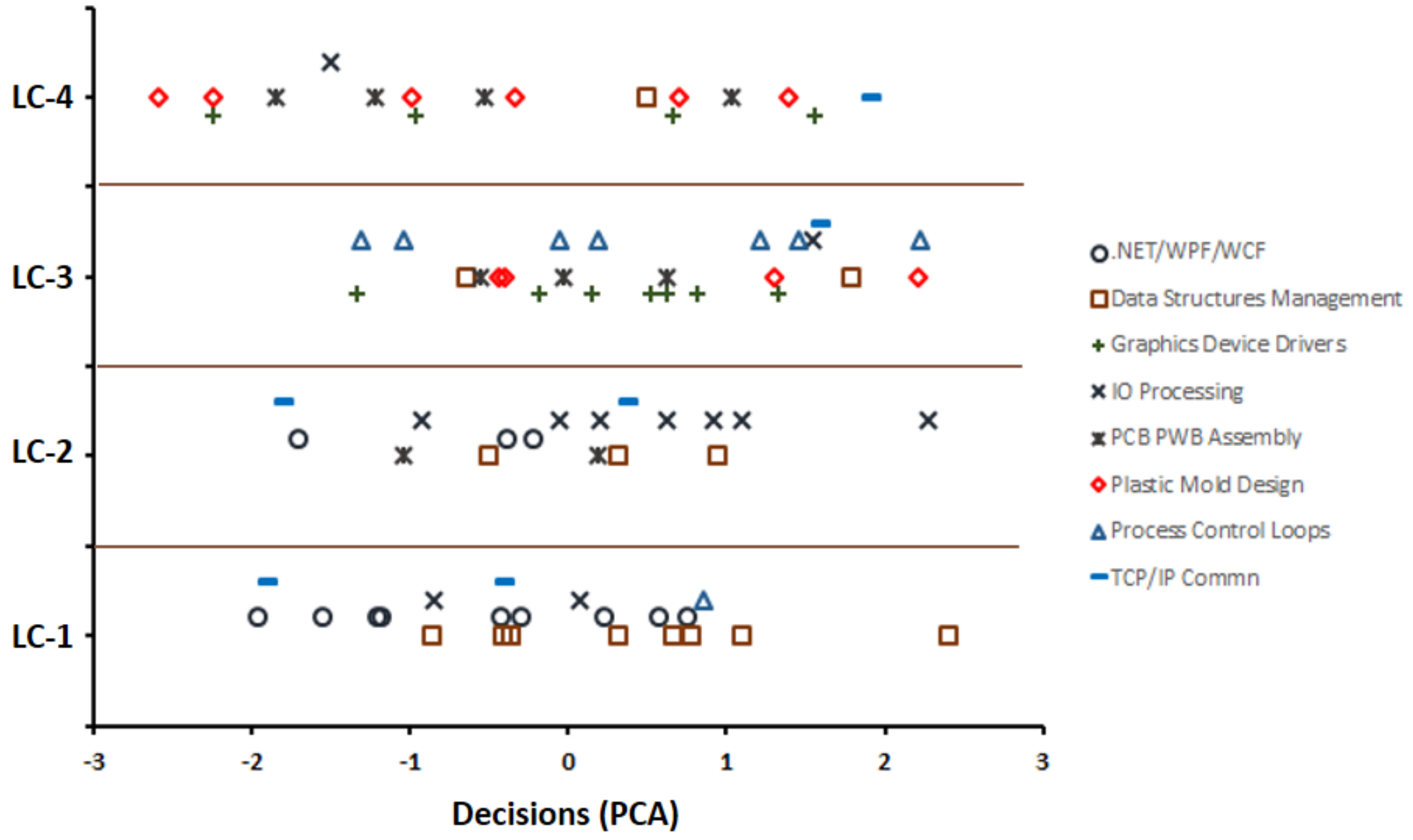

**Figure 20 Analyzing Learning Cycle Consequences**

Design decisions pertaining to various products designed and developed in the organization can thus be codified. As Learning Cycle consequences are experienced for the decision, the same can be captured and codified, as illustrated in Figure 19 . To analyze the Learning Cycle consequence, a visual representation of the different decision points against the corresponding Learning Cycle Consequences is done, as illustrated in Figure 20. The various decision points (each of which corresponds to a point in the multi-dimensional decision space) are reduced to single dimensional values utilizing techniques such as Principal Component Analysis (Martinez and Kak, 2001). The design decision making fidelity can be monitored across the organization, and actions may be taken knowledge area-wise to ensure that progressively, favorable learning cycle consequences (i.e. more of LC-1/LC-2) are increasingly experienced. For instance, the figure indicates that decisions pertaining to "Plastic Mold Design" often encounter re-work. Hence, focused interventions on this knowledge area can be planned, such as inviting external experts, focused deep dive reviews and provisioning of additional risk budget.

### F. *Analyzing Perception-vs-Reality Scenarios*

Figure 21 illustrates the different scenarios based on the perception versus reality situations, and corresponding wastes. The horizontal axis indicates the perceptions of different uncertainty-variability situations, while the vertical axis indicates the corresponding reality. The diagonal represents cases where there is no difference between the perception and the reality, causing most of the product design and development activities to be value added, determining the form of the product through robust design decisions, with minimal wastes. Non-diagonal cases induce moderate or high wastes, due to differences in the perception versus reality.

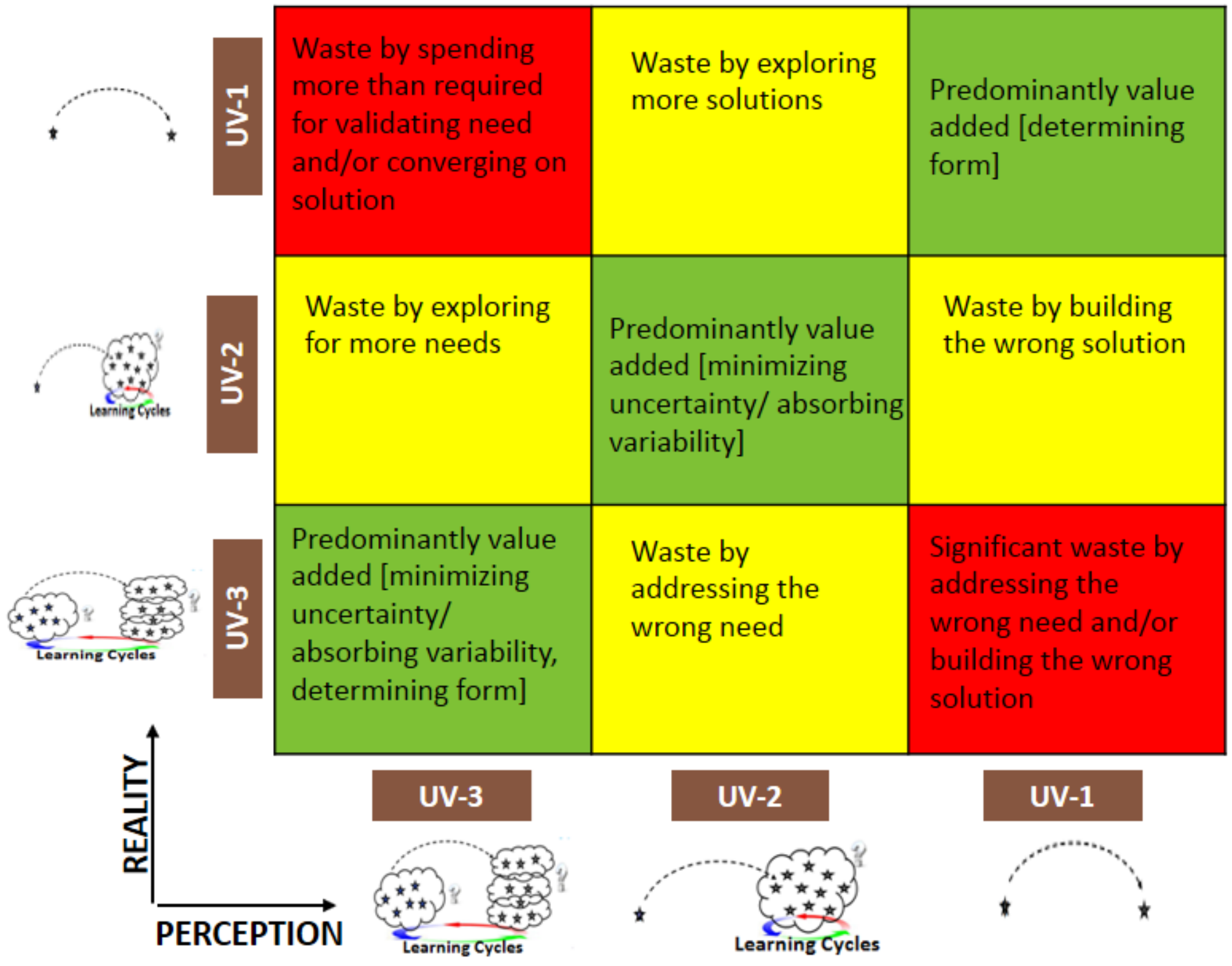

**Figure 21 Perception Reality Scenario Matrix**

For instance, one case is when the design team's perception is of the scenario UV-1, wherein they think they know exactly what the needs are, and the optimal design decisions to be made, but the reality is that of scenario UV-3 of significant uncertainty and variability. As a consequence, significant wastes occur during product design and development, due to team addressing the wrong need, making suboptimal/ wrong decisions that build an inadequate solution, as indicated. Another case is when the design team's perception is of UV-3 where the prevalent knowledge seems inadequate to understand the needs space and arrive at the appropriate solution, whereas the reality is that the knowledge is adequate. This results in the team spending more time in exploring needs and converging on the solution, than required. Actual situations occurring in product design programs can be mapped to one of the cells in the scenario matrix to analyze the predominant scenarios occurring within the organization and take specific actions (such as deep dive reviews and involvement of external experts).

### G. *Uncertainty-Variability Scenarios Driving Evolution of Prevalent Knowledge*

Value creation drives successful innovation, and this value creation surfaces the need for design and development of successful products that add value to customers and various stakeholders. As discussed earlier, optimal and right decisions taken during the product design and development lifecycle result in successful products that stand ahead of the competition. The decisions pertain to different knowledge areas as applicable to the product being designed. The technology maturity corresponding to each knowledge area, and the technology maturity associated with the product serves as a key indicator for the focus areas for successful innovation. Different product life cycle theories are discussed in literature, and applied to analyze the evolution of different product lines – for instance, the case of mobile phones is analyzed in (Funk, 2004), on dimensions such as dominant design and economies of scale. Depending on the appropriate product life cycle theory for the product being designed, mapping the product line evolution stages to the corresponding evolution of prevalent knowledge, can provide additional insights into where the design team needs to focus, towards driving the appropriate level of innovation. S-Curve (Fu et. al., 2010) is one of methodologies used to represent the associated lifecycle of the product, and is well discussed in literature. For instance, four patterns of innovation based on S-Curve are highlighted (Sawaguchi, 2011): "New Technologies" oriented disruptive innovation, "Feature Transfer" oriented disruptive innovation, "New Technologies" oriented sustaining innovation, and "Step by Step" oriented sustaining innovation. In a typical S-Curve, the product lifecycle can be represented into 3 stages - the beginning of the S-curve marks the "birth" of a product addressing a new opportunity, the steep slope represents the growth of the product, and the end of the curve represents the obsolescence or "death" of the product. Usually, the end of one S-curve marks the beginning of a new S-curve for a new product that displaces the earlier product. Figure 22 illustrates the evolution of prevalent knowledge in tandem with the product lifecycle stages. This evolution can be directed based on the position of the product in the S-Curve. The Infancy stage of the product is where the product development team needs to predominantly focus on discovering the appropriate new needs and latent needs in the needs space, and correspondingly explore various solutions in the solution space. In this stage, the product provides some unprecedented function or improvement as compared to existing products in the market. This scenario aligns with UV-3 scenario of Uncertainty and Variability illustrated in Figure 8. Based on this, an efficient knowledge value stream would drive for realizing that it is UV-3, and provide for innovation that is largely radical or exploratory, by flowing across the knowledge across the organization (enhanced knowledge flux) enabling cross pollination. The growth stage in the S-Curve is characterized by significant jump in value and potential, with increasing efficiency and performance of the product. The needs space is well understood, and the focus of the product design team is on significantly enhancing the efficiency, reliability and performance of the product. This scenario aligns with UV-2 scenario of uncertainty and variability. An efficient and effective knowledge value stream provides for innovation that exploits the various choices and options to deliver the stated need. In the maturity stage on S-Curve, the product has already achieved the best performance level, and the corresponding technology has already been developed to the limit. There are no significant knowledge gaps, and the product design does not face any significant challenges, and there is minimal uncertainty and variability (aligning with scenario of UV-1). The focus of innovation shifts towards aspects such as delivery, service and process. This also serves as a trigger for the team to explore initiating a new S-Curve for a paradigm shift for a new product. Thus, by aligning the knowledge value stream with the product life cycle stage, the evolution of prevalent knowledge can be directed towards driving the right level of innovation.

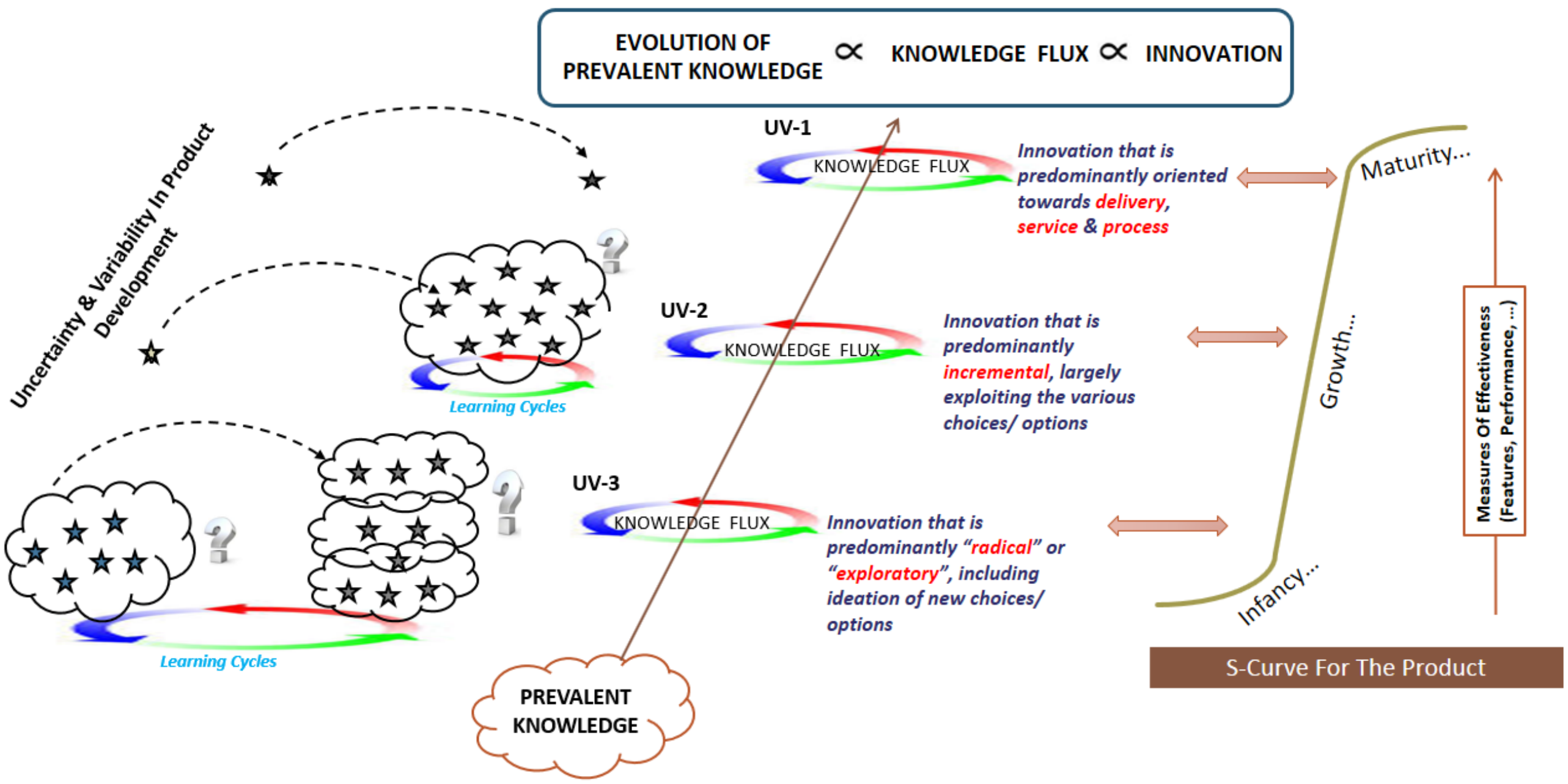

**Figure 22 Directed Evolution of Prevalent Knowledge**

## H. *Elements for Phase-wise Deployment*

A phase-wise approach is required for maturing knowledge value stream, since it involves transformation of culture, process and practice in an organization. Figure 23 illustrates an example of a phase wise maturity model, based on the proposed framework, encompassing both knowledge cadence and learning cycle aspects. Phase 1 involves spreading awareness of knowledge value stream elements and benefits, and surfacing burning issues in making design decisions. Phase 2 comprises baseline maturity assessments of the prevalent processes/ practices for the knowledge value stream elements and knowledge flow and flux. Phase 3 involves specific actions being taken to improve the baseline scores. At the end of phase 3, the required fundamental processes and practices pertaining to knowledge value stream is established. Phase 4 focuses on establishing the closed loop system and accelerating the learning cycles. Phase 5 concentrates on sustenance activities.

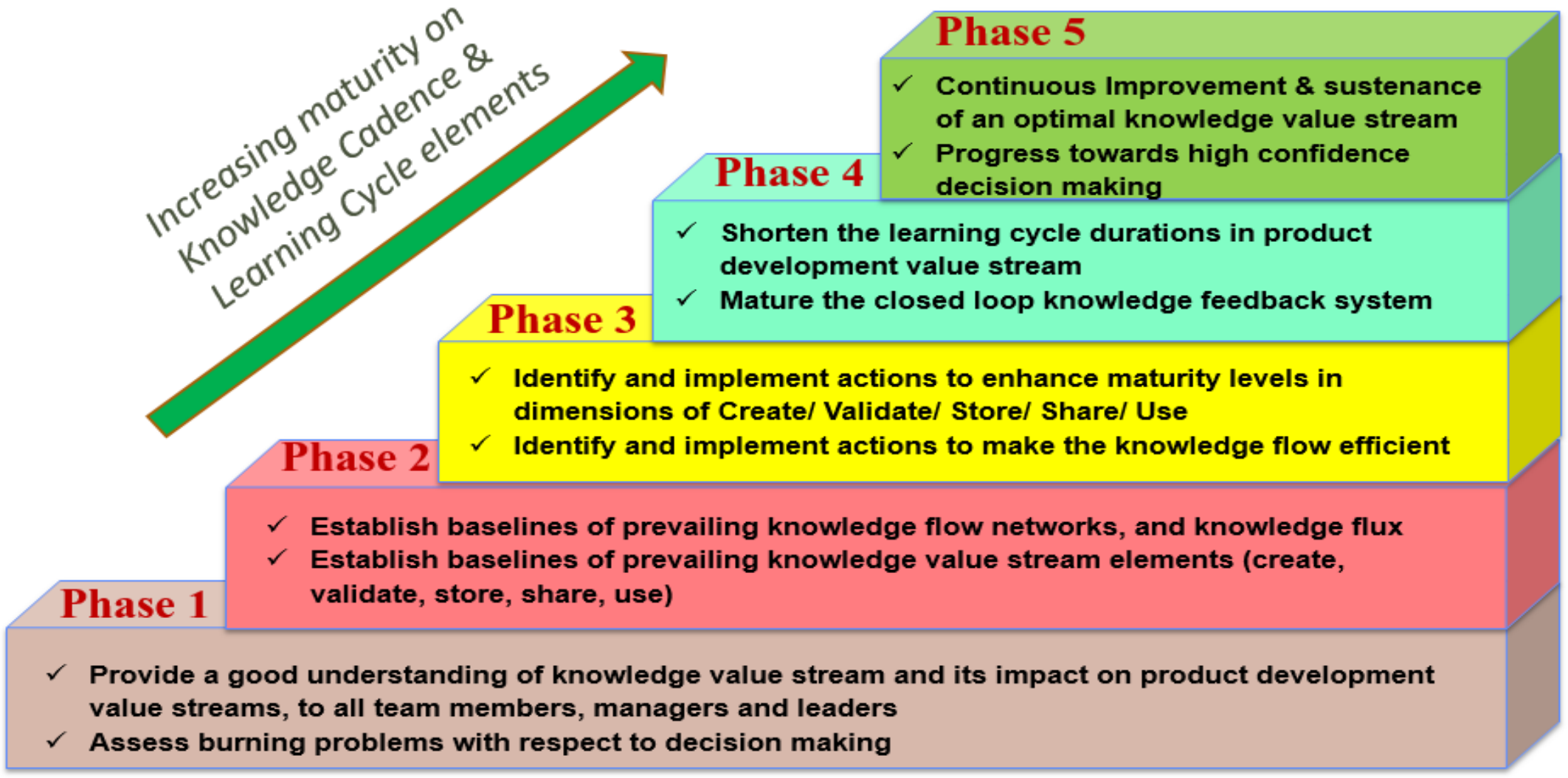

**Figure 23 Phase-wise Deployment**

Figure 24 illustrates a progress maturity pattern based on the phase-wise deployment approach discussed earlier. The initial phases focus on maturity in the CVSS elements. Subsequently, focus on enhancing knowledge flow and knowledge flux elements is recommended. When sufficient levels of maturity are achieved for CVSS, flow and flux, the focus can then be on maturing the closed loop, ensuring favorable learning cycle consequences, and ensuring that perception is not differing from reality.

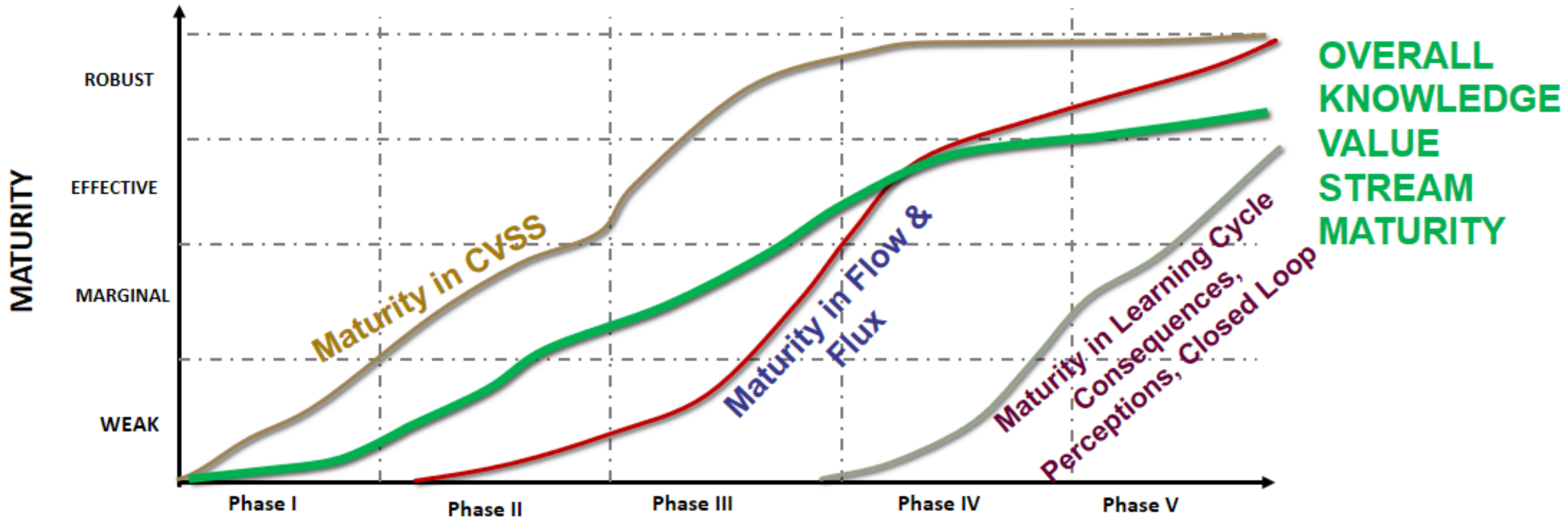

**Figure 24 Phase-wise Maturity of Knowledge Value Stream**

I. *Addressing Wastes in Knowledge Value Stream*

Figure 25 illustrates various points of potential significant wastes in the knowledge value stream.

- Creation: The mix of the form of the prevalent knowledge, i.e. tacit and explicit, has a bearing on wastes that can occur. For instance, if the prevalent knowledge has a high tacit proportion, there are significant chances for knowledge being lost when members move out of the organization (or move to other teams), impacting creation of new knowledge.

- Validation: One level of validation occurs when learning cycles are encountered on the decisions taken. Another level of validation occurs based on reviews by subject matter experts. Wastes can occur when the knowledge gained through learning cycles are not fed back into the knowledge value stream, and/or experts have not vetted the knowledge
- Storage: Potential opportunities for wastes include inadequate storage mechanisms - unorganized, cumbersome navigation, lacking good searching capabilities etc.

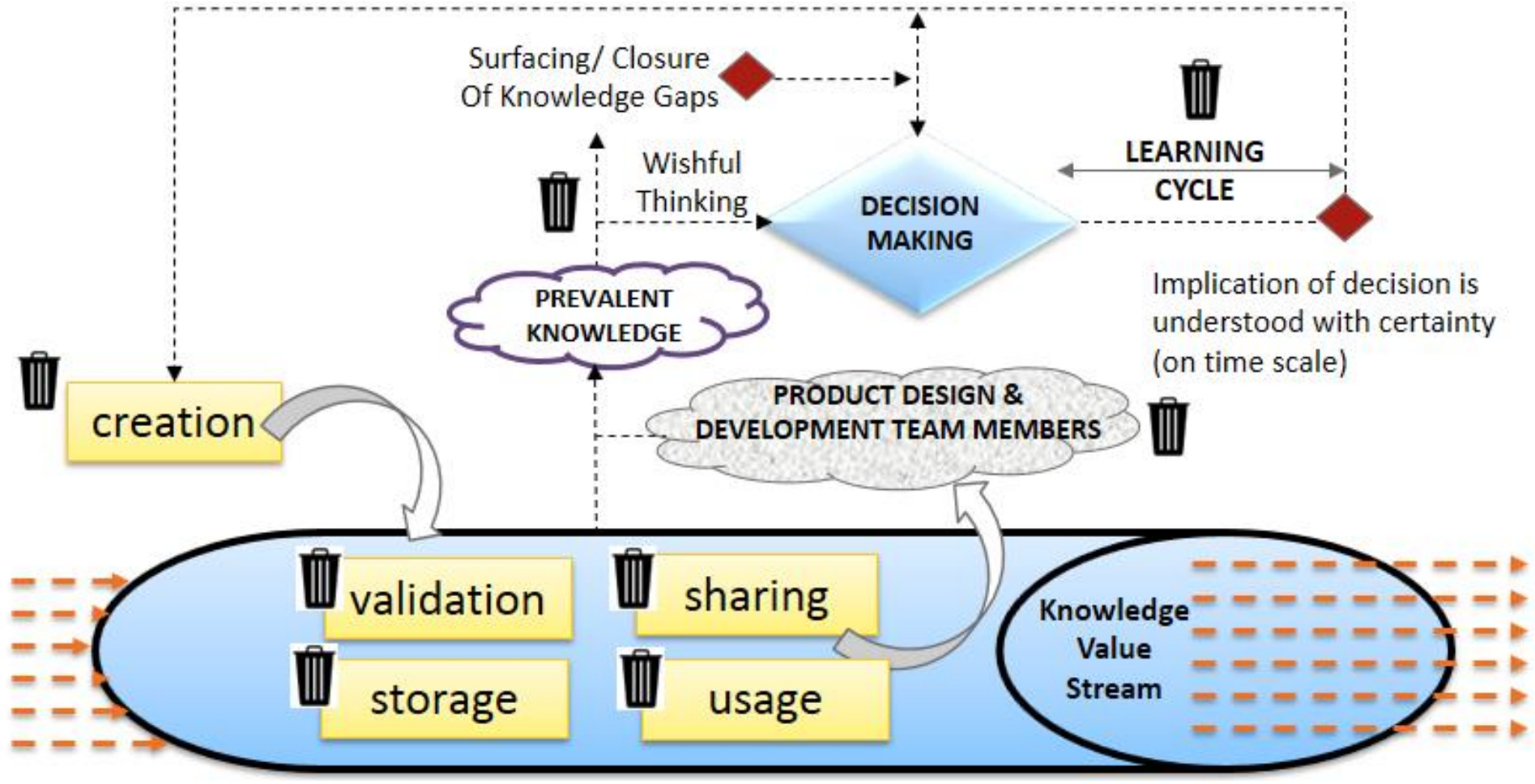

**Figure 25 Points of Waste in Knowledge Value Stream**

- Sharing: Sharing is inhibited when there is knowledge "hoarding" by individuals and groups in the organization. There may be different factors such as trust issues and fear that causes knowledge hoarding by individuals in a team.
- Usage: When knowledge is not in a form that is readily usable, or when there is more time spent on looking for relevant knowledge, wastes occur. Knowledge can be considered usable when it is clear, understandable and actionable.
- Learning Cycle: Long learning cycles with unfavorable consequences (LC-3, LC-4) cause significant amount of rework waste late in the product design and development lifecycle
- Wishful Thinking: When decisions being made without surfacing and closing the knowledge gaps, significant rework wastes occur
- Product design team: Wastes due to differences in perception versus reality cause wastes – illusory progress, or unnecessary activities

The various elements highlighted in the phase-wise deployment of the proposed framework addresses these points of wastes.

## V. Conclusions & Future Work

Knowledge Value Stream focuses on how the knowledge flows to drive the decisions across the product development value streams. For complex products, the design teams face significant challenges in making the right design decisions, due to significant uncertainty, variability, and knowledge gaps. Wrong decisions often result in causing defects, rework loopbacks, cost overruns, product under-performance, and product falling behind competition. This paper proposed a framework for knowledge value stream for architecting and designing complex products, progressively maturing design decisions. The framework has knowledge cadence and learning cycles as its core elements, and incorporates the nuances specific to complex product design and development, including uncertainty, variability, closed feedback loops and perceptions. Systematic assessment of various framework elements, such as knowledge flow, knowledge flux and CVSS maturity, has been illustrated. A phased deployment approach has also been proposed, progressively improving the knowledge value stream maturity. Future work will involve enhancing the proposed framework for handling scenarios pertaining to complex system-of-systems development spanning multiple organizations, taking into consideration knowledge flow dynamics. Formalism of the proposed framework to facilitate analysis for more insights into learning cycles and feedback loops, is another future focus area.